\documentclass[aps,preprint,tightenlines,superscriptaddress,showpacs]{revtex4}
\usepackage{amsmath}
\usepackage{epsfig}
\newcommand{\bra}{\langle}
\newcommand{\ket}{\rangle}

\newcommand{\bs}[1]{\ensuremath{\boldsymbol{#1}}}

\begin{document}

\title{Do $c\bar c n\bar n$ bound states exist?}

\author{J. Vijande}
\affiliation{
Departamento de F\' \i sica Te\'orica e IFIC, 
Universidad de Valencia - CSIC, E-46100 Burjassot, Valencia, Spain}
\affiliation{Departamento de F\'\i sica Fundamental, Universidad de Salamanca, 
E-37008 Salamanca, Spain}
\author{E. Weissman}
\affiliation{The Racah Institute of Physics, The Hebrew University, 91904,
Jerusalem, Israel}
\author{N. Barnea}
\affiliation{The Racah Institute of Physics, The Hebrew University, 91904,
Jerusalem, Israel}
\affiliation{Institute for Nuclear Theory, University of Washington, 
Seattle, WA 98195, USA}
\author{A. Valcarce}
\affiliation{Departamento de F\'\i sica Fundamental, Universidad de Salamanca, E-37008 Salamanca, Spain}

\date{\today}
\begin{abstract}
The four--quark system 
$c\bar c n\bar n$ is studied in the framework of the constituent quark 
model. Using different types 
of quark-quark potentials, we solve the four--body Schr\"odinger equation 
by means of the hyperspherical harmonic formalism. 
Exploring the low laying $J^{PC}$ states for different isospin configurations
no four-quark bound states have been found.
Of particular interest is the possible four-quark
structure of the $X(3872)$. We rule out the possibility that this particle is
a compact tetraquark system, unless additional correlations, either in the
form of diquarks or at the level of the interacting potential, 
not considered in simple quark models do contribute.
\end{abstract}

\pacs{12.39.Jh,14.40.Lb,21.45.+v,31.15.Ja} 
\maketitle

\section{Introduction}

In the last few years the discovery of several new heavy hadrons containing charm quarks has
renewed the interest in heavy quark spectroscopy. In 2003 a new resonance named $X(3872)$ 
was reported by the Belle Collaboration 
in the invariant mass distribution of $J/\psi \pi^+\pi^-$ mesons produced in $B^{\pm}\to K^{\pm} X(3872)\to 
K^{\pm} J/\psi \pi^+\pi^-$ decays. It appeared as a narrow peak with a mass 3871.2 $\pm$ 0.5
MeV and a width $\Gamma <$ 2.3 MeV, consistent with the detector resolution \cite{Bel03}.
This state was confirmed by BaBar \cite{Bab05b},
CDF \cite{CDF04} and D0 Collaborations \cite{D0c05}.
In 2004 Belle reported the observation of a new charmonium state in the 
$\omega J/\psi$ invariant mass distribution for exclusive $B\to K \omega J/\psi$ decays 
\cite{Bel4b}. This state, named $Y(3940)$, has a mass and width of $3943 \pm 11 \pm 13$ MeV
and $87 \pm 22 \pm 26$ MeV, respectively, and it has not been seen in the decay modes 
$Y(3940) \to D \bar D$ and $D\bar D^*$. In July 2005 Belle claimed the observation of a second charmonium 
resonance named $X(3940)$ with a mass of $3943 \pm 6 \pm 6$ MeV and total width of less than $52$ MeV 
\cite{Bel05}. This state has been measured in the $e^+ e^- \to J/\psi X(3940)$ reaction and it has been seen to 
decay to $D\bar D^*$ and not to $\omega J/\psi$ or $D\bar D$. A third state, $Z(3940)$, was reported almost 
simultaneously by Belle in $\gamma\gamma\to D\bar D$ with a mass of $3931 \pm 4 \pm 2$ MeV and a width 
of $20 \pm 8 \pm 3$ MeV \cite{Bel5b}. Being its helicity distribution consistent with $J=2$, an 
identification with the excited $\chi_{c2}$ seems to be natural. Up to now, the last experimental neighbor 
of the charmonium tribe has been reported by BaBar, the $Y(4260)$ \cite{Bab05}. 
This state, with quantum numbers $J^{PC}=1^{--}$, is a broad resonance in the invariant mass spectrum 
of $\pi^+\pi^- J/\psi$ with mass 4259$\pm$8$\pm 4$ MeV and width $83\pm 23 \pm 5$ MeV. 

While some members of this new hadronic zoo may fit in the simple quark model
description as quark-antiquark pairs ($X(3940)$, $Y(3940)$, and $Z(3940)$ may fit
into the $\chi_{c0}$, $\chi_{c1}$, and $\chi_{c2}$ quark model structure)
others appear to be more elusive ($X(3872)$ and $Y(4260)$). With the advent
of all these new resonances, theoretical speculations about the existence of 
four-quark states mixed with $c\bar c$ quark-antiquark bound states have
been reinforced, the $X(3872)$ being the main responsible for that. Before its
discovery, only a few attempts were made to look for $c \bar c n \bar n$ states.
In the early 80's, Gelmini \cite{Gel80} studied the 
$S-$wave $c\bar c n\bar n$ states using the one-gluon-exchange potential 
and virtual annihilation of color pairs, obtaining some candidates that could 
lie below any of the dissociation channels. Chao \cite{Cha80} explored the decay, 
hadronic production, production in $e^+e^-$ annihilation, 
and photoproduction of various types of $c\bar c n\bar n$ states using the quark-gluon model
proposed by Chan and Hogaasen~\cite{Cha77}.
Using a potential derived from the MIT bag model in the
Born-Oppenheimer approximation Chao also concluded that $c\bar c n\bar n$ states lie in the 
3.2$-$3.7 GeV energy range, and therefore below the lowest two-meson thresholds~\cite{Cha81}.
Silvestre-Brac and Semay~\cite{Sil93} analyzed $L=0$ four-quark systems by means of  
the Bhaduri potential through a variational method in a harmonic oscillator basis,
suggesting the existence of several $(Q_1 \bar Q_2 )(n\bar n)$ bound states, among them
the $J^P=0^+$ and $J^P=1^+$ $c\bar c n \bar n$. 
After the discovery of the $X(3872)$ the question about the possible 
existence of $c\bar c n\bar n$ bound states was posed again. Maiani~{\it et al.} \cite{Mai05} 
constructed a model of the $X(3872)$ in terms of diquark-antidiquark
degrees of freedom. Using the $X(3872)$ as input
they predict other $c\bar c n\bar n$ states with quantum 
numbers $0^{++}$, $1^{+-}$, and $2^{++}$. Ebert {\it et al.}~\cite{Ebe05} 
addressed heavy tetraquarks with hidden charm and beauty 
in a diquark-antidiquark relativistic quark model, concluding that the $X(3872)$ 
could be identified with the $1^{++}$ neutral charm tetraquark state. 
The existence of the $X(3872)$ would imply another three four-quark
states close in energy. Hogaasen {\it et al.} \cite{Hog06} explained the 
mass and coupling properties of the $X(3872)$ resonance as a $1^{++}$ 
four-quark state using a chromomagnetic interaction once all 
spin-color configurations compatible with these quantum numbers were included.
Matheus {\it et al.} \cite{Mat07} used QCD spectral sum rules to test the nature of the $X(3872)$ 
within a diquark-antidiquark scheme, identified as a possible $1^{++}$ $c\bar c n\bar n$ candidate.

Analogous alternatives have been scrutinized to interpret the experimental data
of the open-charm meson sector. Several new states with intriguing properties hard
to accommodate in a standard quark-antiquark scheme were reported in recent years.
In 2003 BaBar reported a charm-strange state, the $D_{sJ}^*(2317)$, with a mass of 
2316.8$\pm$0.4$\pm$3 MeV and a width of less than $4.6$ MeV \cite{Bab03}. It was confirmed by CLEO Collaboration 
\cite{Cle03} and also by Belle \cite{Bel04}. Besides, BaBar had also pointed out the existence of another 
charm-strange meson, the $D_{sJ}(2460)$ \cite{Bab03}. This resonance was measured by CLEO \cite{Cle03} and 
confirmed by Belle \cite{Bel04} with a mass of 2457.2$\pm$1.6$\pm$1.3 MeV and a width less than $5.5$ MeV. 
Belle results are consistent with the spin-parity assignments of $J^P=0^+$ for the $D^*_{sJ}(2317)$ and $J^P=1^+$ 
for the $D_{sJ}(2460)$. In the nonstrange sector Belle reported the observation of a nonstrange broad scalar 
resonance, named $D^*_0$, with a mass of $2308\pm 17\pm 15\pm 28$ MeV and a width $276 \pm 21 \pm 18 \pm 60$ MeV 
\cite{Belb4}. A state with similar properties has been suggested by FOCUS Collaboration \cite{Foc04} during 
the measurement of excited charm mesons $D^*_2$.
Although there are several theoretical interpretations for these states (see Ref. \cite{Swa06}),
the difficulties to identify some of them with conventional 
mesons (rather similar to those appearing in the light-scalar meson sector) 
were interpreted as signals indicating that other configurations, for example 
four-quark contributions, could be playing a significant role \cite{Oka06,Vij06}.
The idea behind this interpretation is rather simple.
Physical mesons are easily identified with $q\overline{q}$ 
states when virtual quark loops are not important. This is the case of the pseudoscalar and vector mesons, 
mainly due to the $P-$wave nature of the hadronic dressing. However, in the positive
parity sector it is the $q\overline{q}$ 
pair the one in a $P-$wave state, whereas quark loops may be in a $S-$wave. In this case the intermediate 
hadronic states that are created may play a crucial role in the composition of the resonance, in other words 
unquenching may be important. The vicinity of these components to the lightest $q\bar q$ state implies that 
they have to be considered either as mixed states or compact structures \cite{Jaf07}. 
This has been shown as a possible interpretation of the low-lying light-scalar mesons \cite{Vij05b}. 

As a consequence, the solution of the four--body problem to analyze the contribution
of four-quark components to the meson spectra 
has become recently a basic tool. Most of the approaches 
found in the literature are
variational calculations with different types of trial wave functions. The
rather important interest of this problem requires numerical methods able
to provide with solutions free of numerical uncertainties.
Recently, a new approach based on the hyperspherical formalism was proposed
to solve exactly the four-quark problem \cite{Bar06}. 
The idea is to perform an expansion of the trial wave function in terms of hyperspherical 
harmonic (HH) functions. This allows to generalize the simplicity of the spherical harmonic
expansion for the angular functions of a single particle motion to a system of
particles by introducing a global length $\rho$, called the hyperradius, and a
set of angles, $\Omega$. For the HH expansion to be practical, the
evaluation of the potential energy matrix elements must be feasible. The main
difficulty of this method is to construct HH functions of proper symmetry for a
system of identical particles. This is a difficult problem that may be
overcome by means of the HH formalism based on the symmetrization of the
$N-$body wave function with respect to the symmetric group using the Barnea and
Novoselsky algorithm \cite{Nir9798}. This method, widely used in nuclear physics, 
was applied in Ref. \cite{Bar06} to the analysis of four-charm quark systems. 

In this work we present a study of the $c\bar c n\bar n$ ground 
states using the HH technique.
For this purpose we have generalized the HH formalism of Ref. \cite{Bar06} 
to describe four-quark states of different flavor. 
The manuscript is organized as follows. In Sect. \ref{tech} the procedure
necessary to generalize the 
hyperspherical formalism for studying quark systems of different flavors is described. 
In Sect. \ref{QM} we review two different quark models we will make use of
to test our method and compare with existing results and experiment.
In Sect. \ref{results} we present the results and the analysis of the $c\bar c n\bar n$
spectroscopy. Finally, we summarize in Sect. \ref{summary} our conclusions.

\section{Technical details}
\label{tech}

\subsection{Basis functions}
Within the HH expansion, the four--quark wave function 
can be written as a sum of outer products 
of color, isospin, spin and configuration terms
\begin{equation}
    |\phi_{CISR}\ket= |{\rm Color}\ket |{\rm Isospin}\ket
               \left[|{\rm Spin}\ket \otimes| R \ket \right]^{J M} \, ,
\end{equation}
such that the four-quark state is a color singlet with well defined
parity, isospin and total angular momentum.
In the following we shall present the construction of the basis functions
for the $QQ\bar n \bar n$ and $Q\bar Q n \bar n$ tetraquark systems.
We shall assume that particles $1$ and $2$ are the $Q$-quarks 
and particles $3$ and $4$ are the $n$-quarks.
In the $QQ\bar n \bar n$ case
particles 1 and 2 are identical, and so are 3 and 4.
Consequently, the Pauli principle leads to the following 
conditions,
\begin{equation}\label{pauli}
\hat P_{12}|\phi_{CISR}\ket=\hat P_{34}|\phi_{CISR}\ket=-|\phi_{CISR}\ket \, ,
\end{equation}
$\hat P_{ij}$ being the permutation operator of particles $i$ and $j$.

Coupling the color states of two quarks (antiquarks) can yield two possible
representations, the symmetric $6$-dimensional, $6$ ($\bar 6$),
and the antisymmetric $3$-dimensional, $\bar 3$ ($3$).
Coupling the color states of the quark pair with that of the antiquark pair
must yield a color singlet. Thus, there are only two possible color states for a
$QQ\bar q \bar q$ system \cite{Jaf77},
\begin{equation}
  |{\rm Color}\ket = | C_{12} C_{34}\ket = \{ | \bar 3_{12} 3_{34} \ket , | 6_{12} \bar 6_{34} \ket\}\,.
\end{equation}
These states have well defined symmetry under permutations, Eq. (\ref{pauli}).
Spin states with such symmetry can be obtained coupling the particle spins
in the following way,
\begin{equation}
  |{\rm Spin}\ket = |((s_1,s_2)S_{12},(s_3,s_4)S_{34})S\ket
                  = | (S_{12} S_{34}) S \ket\;.
\end{equation}
The same holds for the isospin, $|{\rm Isospin}\ket=|(i_3,i_4)I_{34} \ket$,
which applies only to the $n$-quarks, thus $I=I_{34}$.

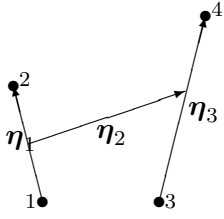
\begin{figure}[t] 
\caption{$H$--type Jacobi vectors.}
\begin{picture}(480,100)(-140,0)
\setlength{\unitlength}{1.1pt}
\put( 60,20){\vector(-1,4){10}}
\put(100,20){\vector(1,4){16}}
\put( 55,40){\vector(3,1){54}} 
\put( 60,20){\circle*{4}}\put(54,18){\scriptsize $1$}
\put( 50,60){\circle*{4}}\put(52,60){\scriptsize $2$}
\put(100,20){\circle*{4}}\put(102,18){\scriptsize $3$}
\put(116,84){\circle*{4}}\put(118,84){\scriptsize $4$}
\put(44,40){ $\bs \eta_1$}
\put(75,42){ $\bs \eta_2$}
\put(110,50){$\bs \eta_3$}
\end{picture}
\label{figH}
\vspace*{-1cm}
\end{figure}

As said, we use the HH expansion to describe the spatial
part of the wave function. We choose for convenience the $H$-type 
Jacobi coordinates (see Fig. \ref{figH}), 
\begin{eqnarray}\label{Jacobi}
\bs{\eta}_1 &=& \mu_{1,2}(\bs r_2 - \bs r_1) \cr
\bs{\eta}_2 &=& \mu_{12,34} 
                 \left(\frac{m_3\bs r_3+m_4\bs r_4}{m_{34}} 
                      -\frac{m_1\bs r_1+m_2\bs r_2}{m_{12}} \right) \cr
\bs{\eta}_3 &=& \mu_{3,4}(\bs r_4 - \bs r_3),
\end{eqnarray}
where $m_{ij}=m_i+m_j$, $\mu_{i,j}=\sqrt{m_i m_j/m_{ij}}$, and $m_{1234}=m_1+m_2+m_3+m_4$.
Using these vectors, it is easy
to obtain basis functions that have well defined symmetry under permutations
of the pairs $(12)$ and $(34)$. The hyperspherical coordinates $(\rho,\Omega)$ 
are defined through the relation
\begin{eqnarray}\label{HS2Jac}
\bs{\eta}_1 &=& \rho \cos{\alpha_2} \cos{\alpha_1} \, \hat{\eta}_1 \cr
\bs{\eta}_2 &=& \rho \cos{\alpha_2} \sin{\alpha_1} \, \hat{\eta}_2 \cr
\bs{\eta}_3 &=& \rho \sin{\alpha_2} \, \hat{\eta}_3 \,,
\end{eqnarray}
where 
\begin{equation}
\rho=\sqrt{\eta_1^2+\eta_2^2+\eta_3^2}\,,
\end{equation}
is the hyperradius, $\hat{\eta}_j \equiv (\theta_j,\, \phi_j)$ is the unit
vector of $\bs{\eta}_j$, 
and $\Omega \equiv (\alpha_1,\alpha_2,\hat{\eta}_1,\hat{\eta}_2,\hat{\eta}_3)$
is a hyperangle that
represents the location on the $8$-dimensional sphere.

By using hyperspherical coordinates one can write the Laplace operator as a sum of two terms
\begin{equation} \label{Laplacen}
  \Delta = \frac{1}{\rho^{8}}\frac{\partial}{\partial \rho}
  {\rho^{8}} \frac{\partial}{\partial \rho}
   - \frac{1}{\rho^2} \hat{K}^2 \; ,
\end{equation}
where the hyperspherical, or grand angular momentum,
operator $\hat{K}^2$ is the $9$ dimensional analogous of the angular
momentum operator associated with the $3$-dimensional Laplacian.

The hyperspherical harmonic functions ${\cal Y}_{[K]}$ are the
eigenfunctions of this hyperangular momentum
operator, labeled by the quantum numbers
$[K]\equiv \{K_3 K_{2} L_3 M_3 L_{2} \ell_3 \ell_2 \ell_1\}$.
The quantum number $K_3$ is the grand hyperangular momentum associated
with the $3$ Jacobi vectors, 
$L_3 M_3$ are the usual orbital angular momentum
quantum numbers of the system, and $\ell_i$ is the angular momentum associated
with the Jacobi vector $\bs{\eta}_i$. The quantum
numbers $K_{2}, L_{2}$ correspond to intermediate coupling of
$\bs{\eta}_1$ and $\bs{\eta}_2$. The explicit expression for the HH functions 
is given by ~\cite{Fab83}
\begin{eqnarray} \label{HH}
 {\cal Y}_{[K]} (\Omega)& = & \big[ \sum_{ m_1, m_2, m_3 }
 \bra \ell_1 m_1  \ell_2 m_2 | L_2 M_2 \ket
 \bra L_{2} M_{2} \ell_3 m_3 | L_3 M_3 \ket 
\cr & & \hspace{-0.7cm}\times
 \prod_{j=1}^{3} Y_{\ell_j, \, m_j} (\hat{\eta}_j) \big] \times
\big[ \prod_{j=2}^{3} {\cal N}_{n_j}^{a_j,b_j}
(\sin \alpha_j )^{\ell_j} (\cos \alpha_j )^{K_{j-1}}
 \cr & & \hspace{-0.7cm} \times
P_{n_j}^{(a_j, b_j)}
(\cos (2\alpha_j) ) \big] \; ,
\end{eqnarray}
where $Y_{\ell, \, m}$ are the spherical harmonic functions, $P_{n}^{(a,b)}$
are the Jacobi polynomials,
$a_j=\ell_j+\frac{1}{2}$, $b_j= K_{j-1} + \frac{3j-5}{2}$
 and ${\cal N}_{n_j}^{a_j, b_j}$ are
normalization constants given by ~\cite{Efr72}:
\begin{equation} \label{norN}
  {\cal N}_{n_j}^{a_j b_j} = \left[ \frac
  {2(2 n_j+a_j+b_j+1) n_j!\Gamma(n_j+a_j+b_j+1)}
  {\Gamma(n_j+a_j+1)
  \Gamma(n_j+b_j+1)}
  \right]^{\frac{1}{2} } \; .
\end{equation}
The quantum numbers $K_j$ are given by
\begin{equation} \label{K_N}
K_j = 2 n_j + K_{j-1}+ \ell_j \,\,\,\, ; \,\,\, n_1 \equiv 0 \; ,
\end{equation}
where the $n_j$'s are non--negative integers. In the following we shall use the
notations $K \equiv K_3$ and $L \equiv L_3$ for the total hyperangular and angular
quantum numbers.
By construction, $\rho^{K} {\cal Y}_{[K]}$ is a harmonic polynomial
of degree $K$, therefore the HH function ${\cal Y}_{[K]}$ is an eigenfunction of
$\hat{K}^2$ with eigenvalues $K(K + 7)$.

The HH functions are a complete basis set for the hyperangular coordinate
$\Omega$. For the hyperradial coordinate our basis functions read
\begin{equation}
   L_{n}(\rho) = \sqrt{\frac{n!}{(n+\nu)!}}
   \rho_0^{-9/2}\left(\frac{\rho}{\rho_0}\right)^{(\nu-8)/2}
   L_{n}^{\nu}\left(\frac{\rho}{\rho_0}\right) e^{-\frac{1}{2}\frac{\rho}{\rho_0}}\,,
\end{equation}
where $L_n^{\nu}(x)$ are the associated Laguerre polynomials. The range
parameter $\rho_0$ and the parameter $\nu$ are varied to get optimal results.

By inspection it can be verified that the spatial basis states, given by 
\begin{equation}
\bra \rho \Omega |R\ket \equiv \bra \rho \Omega |n [K] \ket 
= L_{n}(\rho){\cal Y}_{ [K]} (\Omega)\,,
\end{equation}
 have the following symmetry properties with respect to particle permutations
\begin{equation}
 \hat P_{12} |n [K] \ket = (-1)^{\ell_1}|n [K] \ket \, ; \,
 \hat P_{34} |n [K] \ket = (-1)^{\ell_3}|n [K] \ket
\end{equation}
Application of the Pauli principle, Eq. (\ref{pauli}), to the basis function
\begin{equation}\label{phi_csir}
 |\phi_{CISR}\ket = | C_{12} C_{34} \ket | (S_{12} S_{34}) S \ket 
|I_{34} \ket |n [K] \ket
 \end{equation}
leads to the following restrictions
on the allowed combinations of basis states,
\begin{itemize}
\item[(i)] $(-1)^{S_{12}+\ell_1}=+1$, $(-1)^{S_{34}+I+\ell_3}=-1 \,\,$ 
for the $| 6_{12} \bar 6_{34} \ket$ color state.
\item[(ii)] $(-1)^{S_{12}+\ell_1}=-1$, $(-1)^{S_{34}+I+\ell_3}=+1 \,\,$ 
for the $| \bar 3_{12} 3_{34} \ket$ color state.
\end{itemize}

In the $Q \bar Q  n \bar n$ case particle 2 is the antiparticle of particle 1, 
and particle 4 is the antiparticle of particle 3.
Assuming that $C-$parity is a good symmetry of QCD we can regard quarks and
antiquarks as identical particles and impose the symmetry condition,
Eq. (\ref{pauli}), on the $Q \bar Q  n \bar n$ system as well.
Coupling the color states of a quark and an antiquark can yield two possible
representations: the singlet and the octet.
These representations should be combined in the following way
$\{|1_{12}1_{34}\ket, | 8_{12}, 8_{34}\ket \}$ to yield a total color 
singlet state~\cite{Jaf77}.
However, these states have no definite symmetry under the particle
permutations $(12)$ and $(34)$.
To construct symmetrized states for the $Q\bar Q$ pair we
consider the following combinations,
\begin{equation}\label{defC12}
 | C_{12}^{\Gamma_{12}} \ket = \frac{1}{\sqrt 2} ( | C_{12} \ket + \Gamma_{12}
   | C_{21}\ket  \, ,
\end{equation}
where $C_{12}=\{1_{12},8_{12}\}$, and $\Gamma_{12}=+1$ for a symmetric
combination and $-1$ for antisymmetric. For light quarks the
color and isospin states should be combined together to form states with well
defined symmetry. For $I_z=0$, as an example, these states take the form, 
\begin{equation}\label{defC34}
|( C_{34}\, I_{34})^{\Gamma_{34}}\ket  =  
    +\frac{_1}{^2}\left[|C_{34}\ket\left(|u \bar u\ket\pm |d \bar d\ket\right)
    +\Gamma_{34}|C_{43}\ket\left(|\bar u u\ket \pm |\bar d d\ket\right)\right]\, ,
\end{equation}
where the plus sign stands for $I_{34}=0$ state and the minus sign for the
$I_{34}=1$ state. As before, $C_{34}$ stands for either the singlet or the
octet representation. The total color--isospin states, 
$| C_{12}^{\Gamma_{12}} ( C_{34}\, I_{34})^{\Gamma_{34}} \ket$ are not only
good symmetry states, but also good 
$C-$parity states with,
$C=\Gamma_{12}\Gamma_{34}$. Imposing the Pauli principle for the $Q\bar Q n
\bar n$ system we get the following restrictions,
$\Gamma_{12}(-1)^{S_{12}+\ell_1}=+1$, $\Gamma_{34}(-1)^{S_{34}+\ell_3}=+1$, on
the basis states.

\subsection{Calculation of the matrix elements}

Due to the recursive nature of the HH functions, evaluation of potential matrix
elements become simpler for the pair $(34)$. In this case, the matrix
elements of a isoscalar two-body potential 
\begin{equation}
 V_{ij}=\sum_p V_p(r_{ij}) O_p^S O_p^I O_p^C \,,
\end{equation}
written as a sum of products of spatial, spin, isospin and color operators
respectively,
are diagonal in the quantum numbers $\ell_1,\ell_2,L_2,K_2,S_{12},C_{12},I_{34}$.
The matrix elements of central potentials are further diagonal also in the 
quantum numbers $\ell_3,L,S_{34},S$ thus reducing
the calculation of the matrix elements
\begin{eqnarray} \label{me_csir}
\lefteqn{ \bra \phi_{CISR} | V_{34}| \phi'_{CSIR} \ket
 = \sum_p \bra n K | V_p | n' K' \ket_{K_2 \ell_3}}
\cr &\times&
        \bra S_{34} | O_p^S | S_{34} \ket
        \bra I_{34} | O_p^I | I_{34} \ket 
        \bra C_{34} | O_p^C | C_{34} \ket \,,
\end{eqnarray}
into a product of the internal matrix elements times a two dimensional integral
\begin{eqnarray} \label{int_r}
\lefteqn{  \bra n K | V_p | n' K' \ket_{K_2 \ell_3} = 
\frac{{\cal N}_{n_{3}}^{a,b}{\cal N}_{n'_{3}}^{a,b}}{2^{a+b+2}} 
\int \rho^8 d\rho L_n(\rho) L_{n'}(\rho) }
\cr &\times&
\int_{-1}^{1} dx
(1-x)^a (1+x)^b P_{n_3}^{(a,b)}(x) P_{n'_3}^{(a,b)}(x) V_p(\rho\sqrt{1-x})
\end{eqnarray}
where $x=\cos (2\alpha_2)$, $a=\ell_3+\frac{1}{2}$, and $b= K_2 + 2$. 
Non--central potentials lead to similar expressions after rearrangement
of the angular momentum couplings.

Direct evaluation of the potential matrix elements for other pairs will lead
to multidimensional integrals much more complicated than
Eq. (\ref{int_r}). These complicated integrals can be avoided through
rearrangement of the basis functions. If we denote by $|\phi_{CSIR}(1234)\ket$
the basis function defined in Eq. (\ref{phi_csir}), and by
$|\phi_{CSIR}(ijkl)\ket$ basis function associated with the particles arranged
in the order $ijkl$,
then the interaction matrix element
for the pair $(ij)\neq (34)$ can be evaluated in the following way,
\begin{eqnarray} \label{me_csir_ij}
& & \lefteqn{ \bra \phi_{CISR}(1234) | V_{ij}| \phi'_{CSIR}(1234) \ket =
\sum \bra \phi_{CISR}(1234) | \phi''_{CISR}(klij)\ket} 
\cr & & \times
\bra \phi'''_{CISR}(klij) | \phi'_{CISR}(1234)\ket \times
\bra \phi''_{CISR}(klij) | V_{ij}| \phi'''_{CSIR}(klij) \ket \;.
\end{eqnarray}
The potential matrix element on the rhs of Eq. (\ref{me_csir_ij}) can be
calculated now 
using Eqs. (\ref{me_csir}) and (\ref{int_r}). The permutation matrix elements
are trivial for the isospin and can be calculated using standard $SU(2)$
recoupling techniques for the spin part. The color terms can be obtained for
the $QQ\bar n \bar n$ from the relation
\begin{eqnarray}\label{36to18}
 |1_{13}1_{24} \ket & = &
 \sqrt{\frac{1}{3}} | \bar 3_{12} 3_{34} \ket+\sqrt{\frac{2}{3}}|6_{12}\bar6_{34} \ket  \cr
 |8_{13}8_{24} \ket & = &
-\sqrt{\frac{2}{3}} | \bar 3_{12} 3_{34} \ket+\sqrt{\frac{1}{3}}|6_{12}\bar6_{34} \ket\;,
\end{eqnarray}
and for the $Q \bar Q n \bar n$ from the relations
\begin{eqnarray} \label{18to36}
 |\bar 3_{13}3_{24} \ket & = &
 \sqrt{\frac{1}{3}} | 1_{12} 1_{34} \ket-\sqrt{\frac{2}{3}}|8_{12}8_{34} \ket \cr
 |6_{13}\bar 6_{24} \ket & = &
 \sqrt{\frac{2}{3}} | 1_{12} 1_{34} \ket+\sqrt{\frac{1}{3}}|8_{12}8_{34} \ket \, ;
\end{eqnarray}
and
\begin{eqnarray}\label{18to18}
\label{1_{32}}
 |1_{32}1_{14} \ket & = &
\sqrt{\frac{1}{9}} | 1_{12} 1_{34} \ket+\sqrt{\frac{8}{9}}|8_{12}8_{34} \ket \cr
\label{8_{32}}
 |8_{32}8_{14} \ket & = &
\sqrt{\frac{8}{9}}| 1_{12} 1_{34} \ket-\sqrt{\frac{1}{9}}|8_{12}8_{34} \ket \;.
\end{eqnarray}
In the last case we should also consider the symmetrized states, Eqs. (\ref{defC12}) and
(\ref{defC34}). For these states it can be easily shown that
\begin{eqnarray}
& &  \bra (C I)^{|\Gamma_{12}\Gamma_{34}}(1234)| (C'
  I')^{|\Gamma'_{12}\Gamma'_{34}}(ijkl) \ket 
\cr & &
= \delta_{\Gamma_{12},\Gamma'_{12}} \delta_{\Gamma_{34},\Gamma'_{34}}
  \delta_{I,I'} \bra C(1234)| C'(ijkl) \ket \,.
\end{eqnarray}
Thus reducing the rearrangement matrix elements to Eqs. (\ref{18to36}) and (\ref{18to18}).

The permutation matrix elements for the HH functions are obtained using a
numerical trick due to Efros \cite{Efr95}. Consider the particle positions
$\bs{r}_i$. Using Eqs. (\ref{Jacobi}) and (\ref{HS2Jac}) these positions are
translated into a hyperradius $\rho$ and a point on the hypersphere $\Omega$.
Under particle permutations $(1234)\longrightarrow {(ijkl)}$ 
the point on the hypersphere will move to a new
position, i.e., $\Omega\longrightarrow\Omega_{ijkl}$, 
where the hyperradius remains invariant. 
Using the completeness of the HH
basis and the fact the subspace defined by the quantum numbers $K,L$ is
invariant under particle permutations, we can express the HH functions 
of $\Omega_{ijkl}$ in the following way
\begin{equation}\label{permHHH}
{\cal Y}_{ [K]} (\Omega_{ijkl}) = \sum_{[K']\in KL} 
\bra [K'] | \hat P_{ijkl} | [K] \ket {\cal Y}_{ [K']} (\Omega) \;,
\end{equation}
where $\hat P_{ijkl}$ is the permutation operator.
The sum contains $N_{KL}$ terms, which amounts to the number of HH states with 
hyperspherical
angular momentum $K$ and orbital angular momentum $L$. Eq. (\ref{permHHH}) is
valid for any hyperangular point $\Omega$. Therefore we can choose $N_{KL}$
hyperangular points $\Omega(p),\, p=1\ldots N_{KL}$ and get a set of $N_{KL}$
equations which can be inverted to yield
\begin{equation}\label{permHHT}
\bra [K'] | \hat P_{ijkl} | [K] \ket = 
\sum_{p=1}^{N_{KL}} {\cal Y}_{ [K]} (\Omega_{ijkl}(p)) 
({\cal Y}_{ [K']}(\Omega(p)))^{-1}
 \;.
\end{equation}
Here $({\cal Y}_{ [K']}(\Omega(p)))^{-1}$ stands for the inverse of the
$N_{KL}\times N_{KL}$ matrix
$M$ whose entries are defined as
$M_{[K'],p}={\cal Y}_{ [K']}(\Omega(p))$.

\section{Constituent quark models}
\label{QM}

For our study we will use two standard constituent quark models 
providing a reasonable description of the hadron spectra.
A summary of the energies obtained within both models for selected meson 
states is given in Table \ref{tmeson}, in comparison 
with the corresponding experimental energies.
In the following we draw the basic properties of the interacting potentials. 

\subsection{Bhaduri, Cohler and Nogami model (BCN)}

This model was proposed in the early 80's by Bhaduri {\it et al.} in an
attempt to obtain a  
unified description of meson and baryon spectroscopy \cite{Bha81}. It was
later on applied to  
study the baryon spectra \cite{Sil85} and four-quark ($q q \bar q \bar q$)
systems \cite{Sil93}.  
The model retains the 
most important terms of the one-gluon exchange interaction proposed by de
R\'ujula {\it et al.} \cite{Ruj75},  
namely coulomb and spin-spin terms, and a linear confining potential, having
the form 
\begin{equation}
{V(\vec r_{ij})=-\frac{3}{16}(\vec\lambda^c_i \cdot \vec\lambda^c_j)}
\times \left(\frac{r_{ij}}{a^2}-\frac{\kappa}{r_{ij}}-D+
\frac{\kappa}{m_im_j}\frac{e^{-r_{ij}/r_0}}{r_{ij}r_0^2}(\vec\sigma_i \cdot \vec\sigma_j)\right)\,,
\end{equation}
where $\vec \sigma_i$ are the Pauli matrices and $\vec \lambda^c_i$ are the $SU(3)$ color matrices. 
The parameters
$\kappa=102.67$ MeV fm, $D$=913.5 MeV, $a=$0.0326 MeV$^{-1/2}$ fm$^{1/2}$, $r_0=2.2$ fm, 
$m_{u,d}=$337 MeV, and $m_c=1870$ MeV are taken from Ref. \cite{Sil93}.

\subsection{Constituent Quark Cluster model (CQC)}

This model was proposed in the early 90's in an attempt to
obtain a simultaneous description of the nucleon-nucleon
interaction and the baryon spectra \cite{Rep05}.
It was later on generalized to all flavor sectors giving a reasonable description
of the meson \cite{Vij05a} and baryon spectra \cite{Vij04}. The possible existence of four-quark states
within this model has also been addressed \cite{Vij06,Vij05b,Bar06}. 

The model is based on the assumption that the light-quark 
constituent mass appears because of the spontaneous breaking of the original $SU(3)_{L}\otimes SU(3)_{R}$ 
chiral symmetry at some momentum scale. In this domain of momenta, quarks interact through 
Goldstone boson exchange potentials. 
QCD perturbative effects are taken into account through the one-gluon-exchange (OGE) potential 
as the one used in the BCN model. Finally, it incorporates 
confinement as dictated by unquenched lattice calculations
predicting, for heavy quarks, a screening effect on the 
linearly dependent interquark
potential when increasing the interquark distance \cite{Bal01}.

The model parameters have been taken from Ref. \cite{Vij05a} with 
the exception of the OGE regularization parameter. This 
parameter, taking into account the size of the system, 
was fitted for four--quark states in the description of the light scalar sector \cite{Vij05b}, 
being $\hat r_0=0.18$ fm for mesons and $\hat r_0=0.38$ fm for four-quark systems. 
Let us also notice that the CQC model contains an 
interaction generating flavor mixing
between $n\bar n$ and $s \bar s$ components. It allows to
exactly reproduce the masses of the $\eta$ and $\eta'$ mesons.
In the four--quark case this contribution would mix isospin zero
$Q\bar Q n\bar n$ and $Q\bar Q s \bar s$ components. Such
contributions were explicitly evaluated in the variational
approach of Ref. \cite{Vij05b} for the light isocalar tetraquarks,
giving a negligible effect. Therefore, such a flavor mixing
components will not be consider in the present calculation. In order 
to make a proper comparison between thresholds and four--quark states 
we have recalculated the meson spectra of Ref. \cite{Vij05a} with 
the same $\hat r_0$ value and interaction used in the four-quark 
calculation, neglecting therefore the flavor--mixing terms. The results 
are summarized in Table \ref{tmeson} for the original meson parametrization 
(CQC$_{18}$) and for the one used in this work (CQC) (note that in this 
case the $\eta(547)$ would corresponds to a pure $n\bar n$ state). 
Explicit expressions of the interacting potentials and a more detailed 
discussion of the model can be found in Ref. \cite{Vij05a}.

\section{Results}
\label{results}

\subsection{Threshold determination}

The existence of the color degree of freedom gives rise to an important difference
between four-quark systems and standard baryons or mesons.
For baryons and mesons it is not possible to construct a color singlet
using a subset of the constituents, thus only $q\bar q$ or $qqq$
states are proper solutions of the two- or three-quark
interacting hamiltonian and therefore, all solutions 
correspond to bound states. However, this is not the case for
four-quark systems. The color rearrangement of Eqs. (\ref{36to18})
and (\ref{18to18}),
$(q\bar q)_1\otimes(q\bar q)_1=(qq\bar q\bar q)_1$, makes 
that two isolated mesons are also a solution of the four-quark 
hamiltonian. In order to discriminate between four--quark bound states and
simple pieces of the meson-meson continuum, one has to analyze 
the two-meson states that constitute the
thresholds for each set of quantum numbers.

These thresholds must be determined assuming quantum number conservation 
within exactly the same scheme used in the four-quark calculation. Dealing
with strongly interacting particles, the two-meson states should have well 
defined total angular momentum ($J$), parity ($P$) and $C-$parity ($C$). 
When noncentral forces are not considered, orbital angular 
momentum ($L$) and total spin ($S$) are also good quantum numbers. 
As the systems studied could dissociate either into $(c\bar c)(n \bar n)$
or $(c \bar n) (n \bar c)$, we indicate the lowest two-meson threshold
in both channels, quoting also the final state relative angular momentum. 
We give in Table \ref{th1} the lowest thresholds requiring $J$, $P$, 
and $C$ conservation, while in Tables \ref{th2}, \ref{th3}, 
and \ref{th4} we quote those when $L$ and $S$ are also
preserved. We give the experimental thresholds, 
corresponding to the energies 
in Ref. \cite{PDG06}, the thresholds obtained with 
the BCN model, and those calculated with the CQC model
as described in Sect. \ref{QM}.

A property of $c\bar cn\bar n$ states, that is crucial for the discussion  
on the possible existence of bound states, is that two different physical
thresholds can always be constructed 
for any set of quantum numbers, corresponding
to the $(c \bar c)(n\bar n)$ and $(c\bar n)(n\bar c)$ couplings.
We show an example in Table \ref{th5}, where we give the $J^{PC}=1^{++}$ lowest
threshold in the two possible couplings. 
This is not a general property for any four-quark system, note for instance 
that a $cc\bar n\bar n$ four-quark state only has one allowed physical threshold, corresponding
to the coupling $(c\bar n)(c\bar n)$.

\subsection{The four-quark $c\bar c n\bar n$ spectra}

Once we have developed a method to study four-quark
systems of different flavor, we will address an important
physical question making contact with the actual experimental
situation: Does the quark model naturally predict the
existence of $c\bar c n\bar n$ bound states? 
For this purpose we have performed an exhaustive analysis of the 
$c\bar c n\bar n$ spectra 
by means of the two different quark models, CQC and BCN, 
described in Section \ref{QM}. We have considered all 
isoscalar states with total orbital angular momentum $L\le 1$.
For positive parity, the lowest
states correspond to $L=0$, while $L=1$ for negative parity 
ground states. The reason is that the parity of a 
four-quark state can be written in terms of the relative angular momenta associated 
with the Jacobi coordinates as $P=(-)^{\ell_1+\ell_2+\ell_3}$. 
This makes that $P=-1$ states need three units of relative
angular momentum to obtain $L=0$ ($\ell_1=\ell_2=\ell_3=1$) 
while only one is needed for $L=1$. The same reasoning applies
for $P=+1$ states.
The calculation has been done up to the maximum value of 
$K$ within our computational capabilities, $K_{\rm max}$.

The absolute energy obtained for each state does not 
provide much information regarding the stability of the system.
The relevant quantity is $\Delta_E$, defined as the energy difference 
between the mass of the four-quark system and that of the 
lowest corresponding threshold,
\begin{equation}
\label{delta}
\Delta_E=E_{4q}-E(M_1,M_2)\,.
\end{equation}
Here, $E_{4q}$ stands for the four-quark energy and $E(M_1,M_2)$ for the energy of the 
corresponding threshold. Using this definition, $\Delta_E<0$ will indicate
that all fall--apart strong decays  
are forbidden, i.e., one has a proper bound state. On the other hand, $\Delta_E=0$ 
will indicate that the four-quark solution corresponds to 
an unbound threshold (two free mesons) state.

One of the main difficulties in studying four-quark states was discussed in Ref. \cite{Bar06}, i.e., the 
slow convergence to the asymptotic two free mesons unbound states. This makes
the identification of threshold states
a cumbersome task, demanding large values of $K$. In Ref. \cite{Bar06} it was proposed that 
this problem could be overcome by means of an extrapolation of the four-quark energy using the expression
\begin{equation}
\label{extra}
E(K)=E(K=\infty)+{\frac{a}{K^b}}\,,
\end{equation}
where $E(K=\infty)$, $a$ and $b$ are fitted parameters. 

Although the study of the energy is the most 
powerful way to distinguish between bound and unbound four-quark states we have taken a step 
further analyzing in detail the structure of the wave function. In particular, 
it is possible to determine if a four-quark system behaves as a pure meson-meson state or if it has a 
more involved structure through the analysis of the dominant components of the
wave function. Any solution of 
the four-quark problem that could be identified with a threshold should verify
not only that $\Delta_E=$0 but also 
that the probability of the threshold within the four-quark wave function should be unity.
One can also study the behavior of the root mean square radius (RMS) of the
four-quark system as compared to the radii of the   
two-mesons threshold. The RMS is defined in the usual way for four (two) quark systems
\begin{equation}
RMS|_{4(2)}= \left({\frac{\sum_{i=1}^{4(2)} m_i \langle (\bs{r}_i-
\bs{R}_{CM})^2\rangle}{\sum_{i=1}^{4(2)} m_i}} \right)^{1/2} \, . 
\end{equation}
Combining all this information it could be claimed that any four-quark state
with the following characteristics:
(i) $\Delta_E\to0$ with $K$;
(ii) $RMS\to \infty$ with $K$, (its value should at least exceed the value
corresponding to the threshold system);
(iii) The wave-function tends to a single singlet-singlet
physical channel;
should be considered as an unbound threshold state. 

We present in Table \ref{tC1} the results obtained for all possible $L\le 1$ isoscalar
channels with both quark models, CQC and BCN.
We indicate the maximum value of $K$ used, $K_{\rm max}$. We also indicate the probability of the 
basis vector corresponding to the lowest physical threshold.
Let us first of all concentrate on the results of the two quark models used, 
we will comment later on the comparison with the experimental data.
There is a first general conclusion immediately derived looking at this 
table, namely that no bound state is observed for any set of 
quantum numbers in any of the models, in all cases $\Delta_E>0$. 
Let us analyze the results in detail.
The convergence of the results is illustrated in Table \ref{tC3}, 
where we show the evolution of the energy, radius and probabilities as a function 
of $K$ for two different channels with both quark models.
We have denoted by $P_{C_{12}C_{34}}(S_{12},S_{34})$ 
the probability of the basis vector with color state $C_{12}\otimes C_{34}$ 
and spin $S=S_{12}\otimes S_{34}$ in the 
$(c\bar c)(n\bar n)$ coupling. For each model, we compare in the bottom
part of the table with the lowest energy threshold and its 
RMS (the sum of the RMS's of the two mesons). One can see 
how the four-quark energies do converge to the 
lowest possible threshold at the same time that the radius increases 
linearly and the probability of the vector 
that characterizes the threshold tends to unity. 
When the extrapolation (\ref{extra}) is used,
we observe how the four-quark energies 
reproduce those of the lowest threshold allowed
for each set of quantum numbers. This is illustrated in Table \ref{tC4}, where we 
indicate $E(K=\infty)$ as a function of the initial and final values of $K$ used 
for the fitting. The thresholds are perfectly reproduced within 
a difference, due to the extrapolation, of a few
MeV. These are general features for all the states in Table \ref{tC1}. 

Let us emphasize the importance of comparing the four-quark energies with the
proper mathematical threshold, if this is not done it could easily lead to the
misidentification of bound states. This is made evident on the last columns
of Table \ref{tC1}, where we have quoted 
the experimental thresholds. Looking at column $\Delta_E$ under the 
epigraph Exp., one can see the abundance of {\it spurious}  bound states 
predicted by the BCN model (five) if the experimental thresholds are 
considered instead the correct, theoretical, ones. 

Special attention must also be paid to some numerical approximations
used for solving the four-quark problem.
In the numerical procedure described in Section \ref{tech} 
one can easily restrict the method to perform a calculation
considering only a limited set of relative angular momenta between the quarks.
We show in Table \ref{tlmax} the results obtained neglecting large 
relative orbital angular momenta 
($\sum_i \ell_i\leq1$) for two different set of quantum numbers. 
As can be seen, this approximation is excellent for those
states whose solution can be expressed exclusively in terms 
of $(c\bar c)$ and $(n\bar n)$ mesons 
(the $\ell_i$ are the relative angular momenta 
associated to the Jacobi coordinates 
in the coupling $(c\bar c)(n\bar n)$), as it is the case of  
$J^{PC}=1^{++}$, while it fails for those states with a more 
involved structure. Since one does not know a priori the 
behavior of a particular channel it is inadvisable to perform 
such an approximation and therefore, a full calculation is required 
for any global analysis of four-quark states. 
A similar effect related with the restriction of 
the Hilbert space can also be observed in Ref. \cite{Sil93}, where 
an analysis of the four-quark problem was performed with a variational 
solution of the BCN model in a harmonic oscillator 
basis up to $N=8$,
obtaining 3409 and 3468 MeV for the $J^{PC}=0^{++}$ and $1^{+-}$ states, respectively. 
Being the hyperspherical harmonic basis for $K=8$ 
roughly equivalent to the $N=8$ harmonic oscillator basis, 
we have obtained 3380 and 3436 MeV, respectively, for these states when 
we restrict ourselves to $K=8$. However, once we allow $K$ to take larger values,
the energy decreases more than 200 MeV until the lowest threshold for each channel is obtained.

Among all channels presented in Table \ref{tC1}, three of them deserve
a careful analysis: $(L,S)=(1,0)$ for CQC, and $(L,S)=(1,2)$ for CQC and BCN. 
Their energies, probabilities, and radius are resumed in Tables \ref{taux1} and \ref{taux2} 
as a function of $K$. We observe that they 
have a dominant octet--octet color component that could be 
interpreted as evidence of a compact four--quark system. However, 
since the lowest threshold in all three cases 
corresponds to a $D\bar D$ or $D^*\bar D^*$ state, the coupling where 
the numerical calculation is done, $(c\bar c)(n\bar n)$, 
is not the most appropriate one to identify threshold solutions. 
Once the states are re-expressed in the proper basis, 
$(c\bar n)(n\bar c)$, it can be easily observed how they converge 
to the lowest available threshold as nicely as
the other sets of quantum numbers.

From the discussion above it is clear that no bound state exist 
either for CQC or for BCN. Let us now address the question
if it is possible to generate a bound state by means of reasonable assumptions 
in the interacting hamiltonian, i.e., would a different thoughtful two-body 
quark-quark interaction be able to 
generate a bound state? The answer is that it does not seem to be possible.
The reason for that is simple, whenever one modifies 
the interacting potential, not only the four-quark energy but also the  
two-body energy is affected. This has been illustrated in Table \ref{tevol} for two particular cases,
$(L,S)=(0,0)$ $J^{PC}=0^{++}$ CQC, and
$(L,S)=(1,0)$ BCN. In both cases the relative strength 
of the interactions has been tailored in such a way that the lowest threshold is
modified in a significant quantum number (spin, color,...).
In particular, we have fine tuned the one-gluon exchange interaction by means of the
regularization parameter $r_0$ to modify the threshold from $J/\psi\,\omega\vert_S$ to 
$\eta_c\,\eta\vert_S$ (Table \ref{tevol} upper part) and from $h_c\,\eta\vert_P$ to $D\,\bar D\vert_P$ 
(Table \ref{tevol} lower part).
In both cases we show the values of energies, radius and the relevant probabilities for some values 
of $K$. It can be clearly seen that the alternative sets of parameters do converge to the new lowest threshold.

This behavior makes evident the main difference between $c\bar cn\bar n$ and $c\bar nc \bar n$ systems. 
For the later there is a consensus that there would be stable channels against dissociation into two mesons
if the ratio of the mass of the heavy to the light quark
is large enough \cite{Sil93,Ade82}. One should notice that for these states no other combination than a 
$(c\bar n)(c\bar n)$ two-meson system is allowed for the threshold and therefore, any modification in the 
interaction between the two-charm quarks or the two-light antiquarks, for instance the ratio of the masses, 
would not translate into a modification of the energy of the threshold \cite{Vij07}.

Once it has been observed that no bound state can be obtained by means of any thoughtful two-body potential,
one should reformulate the question we made at the beginning of this section into, Is it possible that 
$c\bar c n\bar n$ bound states naturally exist {\it or other restrictions must be imposed to bind them}?
Two main possibilities have been discussed in the literature in order
 to force four-quark states to be bound. On the one 
hand three- and four-quark interactions that would not be factorizable as a sum of two-body potentials could be 
included in the four-quark hamiltonian. Among these interactions, 
three alternatives have been thoroughly discussed:
color three- and four-quark interactions depending on the color 
$SU(3)$ quadratic and cubic Casimir operators \cite{Dmi01},
flip-flop confining interactions \cite{Mas87}, and string model approaches \cite{Oki05}. 
On the other hand one could choose to directly restrict the Hilbert space in the 
few-body problem, selecting a priori those components 
that may favor the binding of the system, the so-called diquarks \cite{Jaf03}. A diquark (antidiquark) is 
an $S-$wave bound state of two quarks (antiquarks) with particular quantum numbers, i.e., antisymmetric in color, 
flavor, and spin.
This has been explored in recent years not only for four-quark systems, but also for three- and five-quark states 
\cite{Kar06}. Technically, the effect of both hypothesis is the same, migrating probability from the color 
singlet-singlet components to the octet-octet ones. This makes that the two 
asymptotically free mesons are no longer a 
solution of the four-quark hamiltonian and therefore the 
interaction can be tailored to create a compact four-quark 
bound state.

Let us finally notice that when this work was finished Ref. \cite{Vij07b}
has analyzed the stability of $QQ\bar q\bar q$ and
$Q\bar Q q \bar q$ systems in a simple string model considering
only a multiquark confining interaction giving by the minimum
of a flip-flop or a butterfly potential. The ground state of
systems made of two quarks and two antiquaks of equal masses
was found to be below the dissociation threshold. While for
the flavor exotic $QQ\bar q\bar q$ the binding increased when
increasing the mass ratio $m_Q/m_q$, for the cryptoexotic
$Q\bar Q q\bar q$ the effect of symmetry breaking is opposite,
the system being unbound whenever $m_q/m_q >1$. Altough more 
realistic calculations are needed before establishing a
definitive conclusion, the conclusions of Ref. \cite{Vij07b} 
strengtened our result about the doubtful existence of 
$c\bar c n\bar n$ compact states.

\subsection{Isoscalar $J^{PC}=1^{++}$ and $2^{-+}$ quantum numbers and the $X(3872)$}

Since the $X(3872)$ was first reported by Belle in 2003 \cite{Bel03} it has gradually 
become the flagship of a new group of states whose properties make their identification as 
traditional $q\bar q$ states unlikely. In this heterogeneous group we could include states like 
the $Y(2460)$ reported by BaBar, 
and the $D_{sJ}(2317)$ and $D_{sJ}(2460)$ reported by BaBar and CLEO. All these states 
deserve full discussions on their own, however in this section we are going to focus only on the properties of the
aforementioned $X(3872)$.

An average mass of 3871.2$\pm$0.5 MeV and a narrow width of less than 2.3 MeV 
have been reported for the $X(3872)$.
Note the vicinity of this state to the $D^0\overline{D^{*0}}$ threshold, 
$M(D^0\,\overline{D^{*0}})=3871.2\pm1.2$ MeV \cite{PDG06} ($3871.81\pm0.36$ MeV according to 
the last measurement by CLEO \cite{CLE07}). 
With respect to the $X(3872)$ quantum numbers, neither D0 nor BaBar have been able to offer a clear prediction about 
its $J^{PC}$. The isovector nature of this state has been excluded by BaBar due to the 
negative results 
in the search for a charged partner in the decay $B\to X(3872)^-K$, $X(3872)^-\to J/\psi\pi^-\pi^0$
\cite{Bab05c}. CDF has performed a determination of $J^{PC}$ of the $X(3872)$ using 
dipion invariant mass distribution and angular analysis, obtaining that only the assignments $1^{++}$ and 
$2^{-+}$ are able to describe data \cite{CDF06}. 
On the other hand, recent studies by Belle combining angular and kinematic properties 
of the $\pi^+\pi^-$ invariant mass, 
strongly favor a $J^{PC}=1^{++}$ state \cite{Bel05b}, 
and the observation of the $X(3872)\to D^0\overline{D^0}\pi^0$ also 
prefers the $1^{++}$ assignment compared to the $2^{-+}$ \cite{Bel06}. 
Therefore, although some caution is still required until better statistic 
is obtained \cite{Set06}, an isoscalar $J^{PC}=1^{++}$ state seems to 
be the best candidate to describe the properties of the $X(3872)$.
All these properties have triggered intense theoretical speculations about the nature of this state.
Among the possible structures that have been explored one can find tetraquarks, cusps, hybrids, glueballs, and 
molecular states, although in most cases these works have been devoted to the study of a limited set of 
quantum numbers in an attempt to determine the viability of describing its energy together with its width and 
decay modes \cite{Swa06}. 

\begin{figure}
\caption{Energy of the $1^{++}$ state using the CQC (solid line) 
and BCN models (dashed line) as a function of $K$. The insert
in the upper-right corner magnifies the large values of $K$
to show the convergence to the corresponding threshold showed
by a straight line.}
\mbox{\epsfxsize=150mm\epsffile{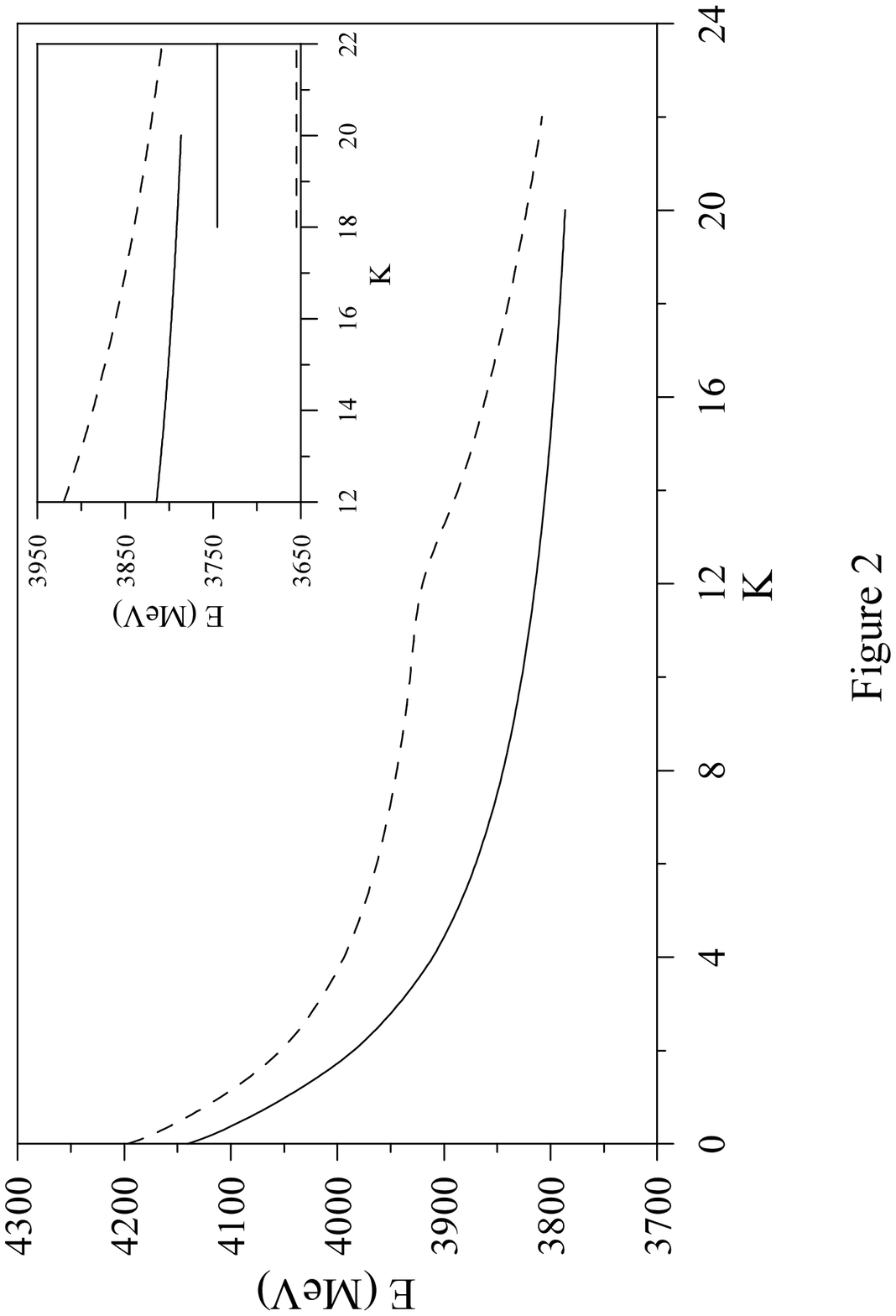}}
\label{fig1}
\end{figure}

Although our main conclusion also applies for these quantum numbers, i.e., the non-existence of four-quark bound states, 
we summarize in detail in this section the results obtained for the isoscalars $1^{++}$ and 
$2^{-+}$ (Table \ref{tX1}) four-quark states. 
In this case we illustrate the convergence plotting the energies
as a function of $K$ in Fig. \ref{fig1}. It can be observed how the BCN $1^{++}$ state does not converge to the lowest 
threshold for small values of $K$, being affected by the presence of an intermediate $J/\psi\,\omega\vert_S$ 
threshold with an energy of 3874 MeV. Once sufficiently large values of $K$ are considered the system follows 
the usual convergence to the lowest threshold (see insert in Fig. \ref{fig1}). 
This behavior can also be observed in the wave function probabilities
(right hand side of Table \ref{tX1}). This is the only 
case where this happens and can be traced back to the unique nature of 
the intermediate threshold, two $S-$wave mesons in a relative $S-$wave. 
Values of $K$ sufficiently large would generate the 
correct solution.

Once all possible quantum numbers of the $X(3872)$ have been analyzed and discarded very 
few alternatives remain. If this state is experimentally proved to be a compact four-quark state
this will point either to the existence of non two-body forces or to the emergence of 
strongly bound diquark structures within the tetraquark. Both possibilities are appealing, does the interaction
becomes more involved with the number of quark or does the Hilbert space becomes simpler? On the one hand, some
Lattice QCD collaborations \cite{Oki05} have reported the important role played by three- and four-quark 
interactions within the confinement (the $Y-$ and $H-$shape). On the other hand, diquark correlations 
have been proposed to play a relevant role in several aspects of QCD, from baryon spectroscopy to
scaling violation \cite{Jaf03}. The spontaneous formation of diquark components can be checked within our formalism.
The four-quark state can be explicitly written in the $(cn)(\bar c\bar n)$ 
coupling in order to isolate the diquark-antidiquark configurations. In the
particular case of the $(L,S)=(0,1)$ $J^{PC}=1^{++}$ of all possible components of the wave function
only two of them have the proper quantum numbers to be identified with a diquark, being the total diquark probability 
less than 3\%. Therefore, it is clear that without any further hypothesis two-body 
potentials do not favor the presence of diquarks and any 
description of these states in terms of diquark-antidiquark 
components would be selecting a restricted Hilbert space.

\section{Summary}
\label{summary}

In this work we present for the first time a generalization of the hyperspherical harmonic
formalism to study systems made of quarks and antiquarks of different flavor. We have focused our 
analysis on hidden-charm systems, namely states containing charm quarks and antiquarks 
with charm quantum number equal to zero. This formalism opens 
the door to an exact study of several other multiquark systems 
containing quarks with different masses and/or 
flavors, like the $s\bar c n\bar n$ or 
the $c\bar c s\bar s$, up to now sparsely analyzed in the literature.

We have performed a systematic analysis of all $c\bar c n\bar n$ isoscalar ground
states. This includes positive parity $L=0$ and negative parity $L=1$ systems with $S=$0,1, and 2.
We have used two standard quark models in the literature, both leading to the same conclusions. 
The relevance of a careful analysis of the numerical thresholds together with 
the numerical approximations involved has been emphasized in order to avoid 
misidentification of bound states.

We have not found any compact four--quark bound state for any set of quantum
numbers. We have studied the possibility of  
generating a bound state by means of a modification of the interacting hamiltonian.
We conclude that no refitting of the models would be
able to force a bound state if only two-body color-dependent forces are considered. 
The reason has been traced back to the particular singlet-singlet decomposition 
available to the $c\bar c n\bar n$ states, namely, the possibility of constructing $(c\bar c)( n\bar n)$ or 
$(c\bar n)( n\bar c)$ two-meson states in such a way that any modification of the two-body 
potential in the four-quark problem is automatically translated into the two-meson final state.
Concerning the $X(3872)$ we have explicitly discussed the quantum numbers favored by experiment, 
$1^{++}$ and $2^{-+}$, obtaining that none of them is bound.

It has been said that when you have eliminated the impossible, whatever remains, however improbable, 
must be the truth \cite{COD}. Therefore, the non-existence of bound $c\bar c n\bar n$ states together with the experimental
observation of suggested non-$q\bar q$ states 
like the $X(3872)$, seems to be clearly emphasizing the need of considering 
new structures not based in naive 
two-body interactions, like for example diquarks configurations or few-body 
potentials, in order to improve our understanding of the hadron spectra.

\section{Acknowledgments}
This work has been partially funded by Ministerio de Ciencia y 
Tecnolog\'{\i}a under Contract No. FPA2004-05616, and by 
Junta de Castilla y Le\'{o}n under Contract No. SA016A07.

\begin{center}
\begin{table}[h!!]
\caption{Meson energies (in MeV) obtained with 
the quark models described in Section \ref{QM}. Experimental data (Exp.) 
are taken from Ref. \protect\cite{PDG06}, except for the state denoted by a dagger
that has been taken from Ref. \protect\cite{Belb4}. See text for the meaning
of the different columns.}
\label{tmeson}
\begin{tabular}{|cc|cccc|}
\hline
($L,S,J,I$)&State               & Exp.                  & CQC$_{18}$    & CQC   & BCN   \\
\hline\hline
\multicolumn{6}{|c|}{$n\bar n$}\\
\hline
(0,0,0,1) &$\pi$                & 139.0                 & 139           & 496   & 136\\
(0,0,0,0) &$\eta(547)$          & 547.51$\pm$0.18       & 572           & 772   & 136\\
(0,1,1,1) &$\rho(770)$          & 775.5$\pm$0.4         & 772           & 744   & 777\\
(0,1,1,0) &$\omega(782)$        & 782.65$\pm$0.12       & 691           & 651   & 777\\
(1,0,1,1) &$b_1(1235)$          & 1229.5$\pm$3.2        & 1234          & 1232  & 1118\\
(1,0,1,0) &$h_1(1170)$          & 1170$\pm$20           & 1257          & 1253  & 1118\\
(1,1,0,1) &$a_0(980)$           & 984.7$\pm$1.2         & 1079          & 1269  & 1254\\
(1,1,0,0) &$f_0(600)$           & 400$-$1200            & 648           & 1262  & 1254\\
(1,1,1,1) &$a_1(1269)$          & 1230$\pm$40           & 1221          & 1269  & 1254\\
(1,1,1,0) &$f_1(1285)$          & 1281.8$\pm$0.6        & 1289          & 1262  & 1254\\
(1,1,2,1) &$a_2(1320)$          & 1318.3$\pm$0.6        & 1315          & 1269  & 1254\\
(1,1,2,0) &$f_2(1270)$          & 1275.4$\pm$1.1        & 1298          & 1262  & 1254\\
\hline
\multicolumn{6}{|c|}{$c\bar n$}\\
\hline
(0,0,0,0) &$D$                  & 1864.5$\pm$0.4        & 1883          & 1936          & 1886 \\
(0,1,1,0) &$D^*(2007)$          & 2006.7$\pm$0.4        & 2010          & 2001          & 2020 \\
(1,1,0,0) &$D^*_0$              & 2308.0$\pm$17$\pm$12$^{\dagger}$      & 2465          & 2498          & 2491 \\
(1,0,1,0) &$D_1(2420)$          & 2422.3$\pm$1.3        & 2492          & 2490          & 2455 \\
(1,1,1,0) &$D^*_1(2430)$        & 2427$\pm$40           & 2504          & 2498          & 2491 \\
(1,1,2,0) &$D^*_2(2460)$        & 2461.1$\pm$1.6        & 2496          & 2498          & 2491 \\
\hline
\multicolumn{6}{|c|}{$c\bar c$}\\
\hline
(0,0,0,0) &$\eta_c(1S)$         & 2980.4$\pm$1.2        & 2990          & 3032  & 3038 \\
(0,1,1,0) &$J/\psi(1S)$         & 3096.916$\pm$0.011    & 3097          & 3094  & 3097\\
(1,0,1,0) &$h_c(1P)$            & 3525.93$\pm$0.27      & 3507          & 3506  & 3502\\
(1,1,0,0) &$\chi_{c0}(1P)$      & 3414.76$\pm$0.35      & 3443          & 3509  & 3519\\
(1,1,1,0) &$\chi_{c1}(1P)$      & 3510.66$\pm$0.07      & 3496          & 3509  & 3519\\
(1,1,2,0) &$\chi_{c2}(1P)$      & 3556.20$\pm$0.09      & 3525          & 3509  & 3519\\
\hline
\end{tabular}
\end{table}
\end{center}

\begin{center}
\begin{table}
\caption{Lowest two-meson thresholds requiring $J$, $P$, and $C$ quantum number conservation. 
Energies are in MeV.}
\label{th1}
\begin{tabular}{|c|cccccccc|}
\hline
        &\multicolumn{2}{c}{Experiment}&&\multicolumn{2}{c}{CQC}&&\multicolumn{2}{c|}{BCN}\\
\hline
$J^{PC}$& $I=0$                 & $I=1$ &$\,\,\,\,$&$I=0$               & $I=1$ &$\,\,\,\,$&$I=0$               & $I=1$ \\
\hline
\hline
$0^{++}$& $\eta_c\,\eta\vert_S$ & $\eta_c\,\pi\vert_S$  & & $J/\psi\,\omega\vert_{S,D}$ & $\eta_c\,\pi\vert_S$ &
        & $\eta_c\,\eta\vert_S$ & $\eta_c\,\pi\vert_S$\\
        & 3528                  & 3119                  & & 3745                        & 3528 & 
        & 3174                  & 3174\\
\hline
$0^{+-}$& $J/\psi f_0\vert_P$   &$h_c\,\pi\vert_P$      & & $\chi_{cJ}\,\omega\vert_P$  & $h_c\,\pi\vert_P$     &       
        & $h_c\,\eta\vert_P$            & $h_c\,\pi\vert_P$\\
        & 3697                  & 3665                  & & 4160                        & 4002 &
        & 3638                  & 3638\\
\hline
$1^{++}$& $\eta_c\, f_0\vert_P$ & $\chi_{c0}\,\pi\vert_P$ & & $J/\psi\,\omega\vert_{S,D}$& $J/\psi\,\rho\vert_{S,D}$    &
        & $\chi_{cJ}\,\eta \vert_P$ & $\chi_{cJ}\,\pi \vert_P$\\
        & 3580                  & 3554                    & & 3745                       & 3838                         & 
        & 3655                  & 3655 \\
\hline
$1^{+-}$& $J/\psi\,\eta\vert_{S,D}$ & $J/\psi\,\pi\vert_{S,D}$ & & $\eta_c\,\omega\vert_{S,D}$  & $J/\psi\,\pi\vert_{S,D}$ &
        & $J/\psi\,\eta\vert_{S,D}$ & $J/\psi\,\pi\vert_{S,D}$ \\ 
        & 3644                      & 3236                     & & 3683                         & 3590                     &
        & 3233                      & 3233\\
\hline
$2^{++}$&$\eta_c\,\eta\vert_D$  &$\eta_c\,\pi\vert_D$   & & $J/\psi\,\omega\vert_{S,D}$ & $\eta_c\,\pi\vert_D$  & 
        & $\eta_c\,\eta \vert_D$& $\eta_c\,\pi\vert_D$ \\
        & 3528                  & 3119                  & & 3745                        & 3528                  & 
        & 3174                  & 3174 \\
\hline
$2^{+-}$&$J/\psi\,\eta\vert_D$  &$J/\psi\,\pi\vert_D$   & & $\eta_c\,\omega\vert_D$     & $J/\psi\,\pi\vert_D$  & 
        & $J/\psi\,\eta\vert_D$ & $J/\psi\,\pi\vert_D$\\ 
        & 3644                  & 3236                  & & 3683                        & 3590                  &
        & 3233                  & 3233\\
\hline
\hline
$0^{-+}$&$\eta_c\,f_0\vert_S$   & $\chi_{c0}\,\pi\vert_{S,D}$   & & $J/\psi\,\omega\vert_P$     & $J/\psi\,\rho\vert_P$&
        &$\chi_{cJ}\,\eta\vert_{S,D}$ & $\chi_{cJ}\,\pi\vert_{S,D}$\\
        & 3580                  & 3554                          & & 3745                        & 3838                  &
        & 3655                  & 3655\\
\hline
$0^{--}$& $J/\psi\,\eta\vert_P$ & $J/\psi\,\pi\vert_P$  & & $\eta_c\,\omega\vert_P$     & $J/\psi\,\pi\vert_P$  &
        & $J/\psi\,\eta\vert_P$ & $J/\psi\,\pi\vert_P$\\
        & 3644                  & 3236                  & & 3683                        & 3590                  &
        & 3233                  & 3233\\
\hline
$1^{-+}$&$\eta_c\,\eta\vert_P$  &$\eta_c\,\pi\vert_P$   & & $J/\psi\,\omega\vert_P$     &$\eta_c\,\pi\vert_P$   &
        & $\eta_c\,\eta \vert_P$& $\eta_c\,\pi \vert_P$\\
        & 3528                  & 3119                  & & 3745                        & 3528                  &
        & 3174                  & 3174\\
\hline
$1^{--}$&$J/\psi\,\eta\vert_P$  &$J/\psi\,\pi\vert_P$   & & $\eta_c\,\omega\vert_P$     &$J/\psi\,\pi\vert_P$   &
        & $J/\psi\,\eta\vert_P$ & $J/\psi\,\pi\vert_P$\\ 
        & 3644                  & 3236                  & & 3683                        & 3590                  &
        & 3233                  & 3233\\
\hline
$2^{-+}$&$\eta_c\,f_0\vert_D$   &$\chi_{c0}\,\pi\vert_D$& & $J/\psi\,\omega\vert_P$     & $J/\psi\,\rho\vert_P$ & 
        & $\chi_{cJ}\,\eta\vert_{S,D}$ & $\chi_{cJ}\,\pi\vert_{S,D}$\\
        & 3580                  & 3554                  & & 3745                        & 3838                  &
        & 3655                  & 3655 \\
\hline
$2^{--}$&$J/\psi\,\eta\vert_P$  & $J/\psi\,\pi\vert_P$  & & $\eta_c\,\omega\vert_P$     & $J/\psi\,\pi\vert_P$  & 
        & $J/\psi\,\eta\vert_P$ & $J/\psi\,\pi\vert_P$\\ 
        & 3644                  & 3236                  & & 3683                        & 3590                  & 
        & 3233 & 3233\\
\hline
$3^{-+}$&$J/\psi\,\omega\vert_P$& $\chi_{c1}\,\pi\vert_D$& & $J/\psi\,\omega\vert_P$    & $J/\psi\,\rho\vert_P$ &
        & $\chi_{cJ}\,\eta\vert_D$ & $\chi_{cJ}\,\pi\vert_D$\\
        & 3880                  & 3650                   & & 3745                       & 3838                  &
        & 3655                  & 3655 \\
\hline
$3^{--}$&$J/\psi\,f_0\vert_D$   &$h_c\,\pi\vert_D$      & & $D^*\,\bar D^*\vert_P$      & $D^*\,\bar D^*\vert_P$ &
        & $h_c\,\eta\vert_D$ & $h_c\,\pi\vert_D$ \\
        & 3697                  & 3665                  & & 4002                        & 4002                  &
        &3638                   & 3638 \\
\hline
\end{tabular}
\end{table}
\end{center}

\begin{center}
\begin{table}[htb]
\caption{Lowest two-meson experimental thresholds imposing 
$L$, $S$, $J$, $P$, and $C$ quantum number conservation. Energies are in MeV}
\label{th2}
\begin{tabular}{|c||c|c|c||c|c|c|}
\hline
        &\multicolumn{6}{c|}{Experiment}\\
\cline{2-7}
$J^{PC}$&\multicolumn{3}{c||}{$I=0$}    & \multicolumn{3}{c|}{$I=1$} \\
\cline{2-7}
        &$(L,S)=(0,0)$          &$(L,S)=(0,1)$  &$(L,S)=(0,2)$  &$(L,S)=(0,0)$  &$(L,S)=(0,1)$  &$(L,S)=(0,2)$  \\
\hline
$0^{++}$& $\eta_c\,\eta\vert_S$ &$-$            &$-$            & $\eta_c\,\pi\vert_S$  &$-$            &$-$ \\
        & 3528                  &$-$            &$-$            & 3119                  &$-$            &$-$ \\
\hline
$0^{+-}$& $J/\psi f_0\vert_P$   &$-$            &$-$            & $h_c\,\pi\vert_P$     &$-$            &$-$ \\
        & 3697                  &$-$            &$-$            & 3665                  &$-$            &$-$    \\      
\hline
$1^{++}$& $-$                   & $\eta_c\, f_0\vert_P$ & $-$   & $-$                   & $\chi_{c0}\,\pi\vert_P$ &$-$\\ 
        & $-$                   & 3580                  & $-$   & $-$                   & 3554 & $-$ \\
\hline
$1^{+-}$& $-$                   & $J/\psi\,\eta\vert_S$ &$-$    & $-$                   & $J/\psi\,\pi\vert_S$  &$-$\\ 
        & $-$                   & 3644                  & $-$   & $-$                   & 3236 &$-$\\
\hline
$2^{++}$& $-$           &$-$            &$J/\psi\,\omega\vert_S$& $-$           &$-$            &$J/\psi\,\rho\vert_S$ \\
        & $-$           &$-$            &3880                   & $-$           &$-$            & 3873 \\
\hline
$2^{+-}$& $-$           &$-$            &$J/\psi\,f_0\vert_P$   & $-$           &$-$            & $J/\psi\,a_0\vert_P$\\ 
        & $-$           &$-$            &3697                   & $-$           &$-$            & 4082 \\
\hline
\hline
        &$(L,S)=(1,0)$          &$(L,S)=(1,1)$  &$(L,S)=(1,2)$  &$(L,S)=(1,0)$  &$(L,S)=(1,1)$  &$(L,S)=(1,2)$  \\
\hline
$0^{-+}$& $-$                   &$\eta_c\,f_0\vert_S$   & $-$                           & $-$   
        &  $\chi_{c0}\,\pi\vert_S$      &$-$    \\
        & $-$                   & 3580                  & $-$                           & $-$   
        & 3554                          & $-$\\
\hline
$0^{--}$& $-$                   & $J/\psi\,\eta\vert_P$ & $-$                           & $-$   
        & $J/\psi\,\pi\vert_P$          &$-$    \\
        & $-$                   & 3644                  & $-$                           & $-$   
        & 3236                          & $-$           \\
\hline
$1^{-+}$& $\eta_c\,\eta\vert_P$ & $D\,\bar D^*\vert_P$  & $J/\psi\,\omega\vert_P$       & $\eta_c\,\pi\vert_P$  
        & $\chi_{c1}\,\pi\vert_{S,D}$   & $J/\psi\,\rho\vert_P$\\
        & 3528                  & 3871                  & 3880                          & 3119                  
        & 3650                          & 3873  \\
\hline
$1^{--}$&$J/\psi\,f_0\vert_{S,D}$& $J/\psi\,\eta\vert_P$& $J/\psi\,f_0\vert_{S,D}$      & $h_c\,\pi\vert_{S,D}$ 
        & $J/\psi\,\pi\vert_P$          & $D^*\,\bar D^*\vert_P$        \\
        & 3697                   & 3644                 & 3697                          & 3665                  
        & 3236                          & 4018  \\
\hline
$2^{-+}$& $-$                   & $\eta_c\,f_0\vert_D$  & $J/\psi\,\omega\vert_P$       & $-$   
        &  $\chi_{c0}\,\pi\vert_S$      &$J/\psi\,\rho\vert_P$  \\
        & $-$                   & 3580                  &3880                           & $-$   
        & 3554                          & 3873\\
\hline
$2^{--}$&$-$                    & $J/\psi\,\eta\vert_P$ & $J/\psi\,f_0\vert_D$          & $-$   
        & $J/\psi\,\pi\vert_P$          &$D^*\,\bar D^*\vert_P$ \\
        & $-$                   & 3644                  & 3697                          & $-$   
        & 3236                          & 4018          \\
\hline
$3^{-+}$& $-$                   &$-$                    & $J/\psi\,\omega\vert_P$       & $-$   
        & $-$                           &$J/\psi\,\rho\vert_P$  \\
        & $-$                   & $-$                   & 3880                          & $-$   
        & $-$                           & 3873\\
\hline
$3^{--}$& $-$                   & $-$                   &$J/\psi\,f_0\vert_D$           & $-$   
        & $-$                           &$D^*\,\bar D^*\vert_P$ \\
        & $-$                   & $-$                   & 3697                          & $-$   
        & $-$                           & 4018          \\
\hline
\end{tabular}
\end{table}
\end{center}

\begin{center}
\begin{table}[htb]
\caption{Same as Table \ref{th2} for CQC.}
\label{th3}
\begin{tabular}{|c||c|c|c||c|c|c|}
\hline
        &\multicolumn{6}{c|}{CQC}\\
\cline{2-7}
$J^{PC}$&\multicolumn{3}{c||}{$I=0$}    & \multicolumn{3}{c|}{$I=1$} \\
\cline{2-7}
        &$(L,S)=(0,0)$          &$(L,S)=(0,1)$  &$(L,S)=(0,2)$  &$(L,S)=(0,0)$  &$(L,S)=(0,1)$  &$(L,S)=(0,2)$  \\
\hline
$0^{++}$& $J/\psi\,\omega\vert_S$       &$-$            &$-$            & $\eta_c\,\pi\vert_S$  &$-$            &$-$ \\
        & 3745                  &$-$            &$-$            & 3528                  &$-$            &$-$ \\
\hline
$0^{+-}$& $\chi_{cJ}\,\omega\vert_P$    &$-$            &$-$            & $h_c\,\pi\vert_P$     &$-$            &$-$ \\
        & 4160                  &$-$            &$-$            & 4002                  &$-$            &$-$    \\      
\hline
$1^{++}$& $-$                   & $J/\psi\,\omega\vert_S$       & $-$   & $-$           & $J/\psi\,\rho\vert_S$ &$-$\\ 
        & $-$                   & 3745                  & $-$   & $-$                   & 3838 & $-$ \\
\hline
$1^{+-}$& $-$                   & $\eta_c\,\omega\vert_S$       &$-$    & $-$           & $J/\psi\,\pi\vert_S$  &$-$\\ 
        & $-$                   & 3683                  & $-$   & $-$                   & 3590 &$-$\\
\hline
$2^{++}$& $-$           &$-$            &$J/\psi\,\omega\vert_S$& $-$           &$-$            &$J/\psi\,\rho\vert_S$ \\
        & $-$           &$-$            &3745                   & $-$           &$-$            & 3838 \\
\hline
$2^{+-}$& $-$           &$-$            &$\chi_{cJ}\,\omega \vert_P$& $-$               &$-$    & $\chi_{cJ}\,\rho \vert_P$\\ 
        & $-$           &$-$            &4160                   & $-$           &$-$            & 4253 \\
\hline
\hline
        &$(L,S)=(1,0)$          &$(L,S)=(1,1)$  &$(L,S)=(1,2)$  &$(L,S)=(1,0)$  &$(L,S)=(1,1)$  &$(L,S)=(1,2)$  \\
\hline
$0^{-+}$& $-$                   &$J/\psi\,\omega\vert_P$        & $-$                           & $-$   
        &  $J/\psi\,\rho\vert_P$        &$-$    \\
        & $-$                   & 3745                  & $-$                           & $-$   
        & 3838                          & $-$\\
\hline
$0^{--}$& $-$                   & $\eta_c\,\omega\vert_P$       & $-$                           & $-$   
        & $J/\psi\,\pi\vert_P$          &$-$    \\
        & $-$                   & 3683                  & $-$                           & $-$   
        & 3590                          & $-$           \\
\hline
$1^{-+}$& $J/\psi\,\omega\vert_P$       & $J/\psi\,\omega\vert_P$       & $J/\psi\,\omega\vert_P$       & $\eta_c\,\pi\vert_P$  
        & $J/\psi\,\rho\vert_P$ & $J/\psi\,\rho\vert_P$\\
        & 3745                  & 3745                  & 3745                          & 3528                  
        & 3838                          & 3838  \\
\hline
$1^{--}$& $D\,\bar D\vert_P$            & $\eta_c\,\omega\vert_P$& $D^*\,\bar D^*\vert_P$       & $D\,\bar D\vert_P$
        & $J/\psi\,\pi\vert_P$          & $D^*\,\bar D^*\vert_P$        \\
        & 3872                   & 3683                 & 4002                          & 3872                  
        & 3590                          & 4002  \\      
\hline
$2^{-+}$& $-$                   & $J/\psi\,\omega\vert_P$       & $J/\psi\,\omega\vert_P$       & $-$   
        &  $J/\psi\,\rho\vert_P$        &$J/\psi\,\rho\vert_P$  \\
        & $-$                   & 3745                  &3745                           & $-$   
        & 3838                          & 3838\\
\hline
$2^{--}$&$-$                    & $\eta_c\,\omega\vert_P$       & $D^*\,\bar D^*\vert_P$                & $-$   
        & $J/\psi\,\pi\vert_P$          &$D^*\,\bar D^*\vert_P$ \\
        & $-$                   & 3683                  & 4002                          & $-$   
        & 3590                          & 4002          \\
\hline
$3^{-+}$& $-$                   &$-$                    & $J/\psi\,\omega\vert_P$       & $-$   
        & $-$                           &$J/\psi\,\rho\vert_P$  \\
        & $-$                   & $-$                   & 3745                          & $-$   
        & $-$                           & 3838\\
\hline
$3^{--}$& $-$                   & $-$                   &$D^*\,\bar D^*\vert_P$         & $-$   
        & $-$                           &$D^*\,\bar D^*\vert_P$ \\
        & $-$                   & $-$                   & 4002                          & $-$   
        & $-$                           & 4002          \\
\hline
\end{tabular}
\end{table}
\end{center}

\begin{center}
\begin{table}[htb]
\caption{Same as Table \ref{th2} for BCN.}
\label{th4}
\begin{tabular}{|c||c|c|c||c|c|c|}
\hline
        &\multicolumn{6}{c|}{BCN}\\
\cline{2-7}
$J^{PC}$&\multicolumn{3}{c||}{$I=0$}    & \multicolumn{3}{c|}{$I=1$} \\
\cline{2-7}
        &$(L,S)=(0,0)$          &$(L,S)=(0,1)$  &$(L,S)=(0,2)$  &$(L,S)=(0,0)$  &$(L,S)=(0,1)$  &$(L,S)=(0,2)$  \\
\hline
$0^{++}$& $\eta_c\,\eta\vert_S$ &$-$            &$-$            & $\eta_c\,\pi\vert_S$  &$-$            &$-$ \\
        & 3174                  &$-$            &$-$            & 3174                  &$-$            &$-$ \\
\hline
$0^{+-}$& $h_c\,\eta\vert_P$    &$-$            &$-$            & $h_c\,\pi\vert_P$     &$-$            &$-$ \\
        & 3638                  &$-$            &$-$            & 3638                  &$-$            &$-$    \\      
\hline
$1^{++}$& $-$                   & $\chi_{cJ}\,\eta \vert_P$     & $-$   & $-$           & $\chi_{cJ}\,\pi \vert_P$ &$-$\\ 
        & $-$                   & 3655                  & $-$   & $-$                   & 3655 & $-$ \\
\hline
$1^{+-}$& $-$                   & $J/\psi\,\eta\vert_S$ &$-$    & $-$           & $J/\psi\,\pi\vert_S$  &$-$\\ 
        & $-$                   & 3233                  & $-$   & $-$                   & 3233 &$-$\\
\hline
$2^{++}$& $-$           &$-$            &$J/\psi\,\omega\vert_S$& $-$           &$-$            &$J/\psi\,\rho\vert_S$ \\
        & $-$           &$-$            &3874                   & $-$           &$-$            & 3874 \\
\hline
$2^{+-}$& $-$           &$-$            &$\chi_{cJ}\,\omega \vert_P$& $-$               &$-$    & $\chi_{cJ}\,\rho \vert_P$\\ 
        & $-$           &$-$            &4296                   & $-$           &$-$            & 4296 \\
\hline
\hline
        &$(L,S)=(1,0)$          &$(L,S)=(1,1)$  &$(L,S)=(1,2)$  &$(L,S)=(1,0)$  &$(L,S)=(1,1)$  &$(L,S)=(1,2)$  \\
\hline
$0^{-+}$& $-$                   &$\chi_{cJ}\,\eta \vert_{S,D}$  & $-$                           & $-$   
        &  $\chi_{cJ}\,\pi \vert_S$     &$-$    \\
        & $-$                   & 3655                  & $-$                           & $-$   
        & 3655                          & $-$\\
\hline
$0^{--}$& $-$                   & $J/\psi\,\eta\vert_P$ & $-$                           & $-$   
        & $J/\psi\,\pi\vert_P$          &$-$    \\
        & $-$                   & 3233                  & $-$                           & $-$   
        & 3233                          & $-$           \\
\hline
$1^{-+}$&  $\eta_c\,\eta\vert_P$        & $\chi_{cJ}\,\eta \vert_{S,D}$ & $J/\psi\,\omega\vert_P$       & $\eta_c\,\pi\vert_P$  
        & $\chi_{cJ}\,\pi \vert_{S,D}$  & $J/\psi\,\rho\vert_P$\\
        & 3174                  & 3655                  & 3874                          & 3174                  
        & 3655                          & 3874  \\
\hline
$1^{--}$& $h_c\,\eta\vert_{S,D}$        & $J/\psi\,\eta\vert_P$& $D^*\,\bar D^*\vert_P$ & $h_c\,\pi\vert_{S,D}$
        & $J/\psi\,\pi\vert_P$          & $D^*\,\bar D^*\vert_P$        \\
        & 3638                   & 3233                 & 4040                          & 3638                  
        & 3233                          & 4040  \\      
\hline
$2^{-+}$& $-$                   & $\chi_{cJ}\,\eta \vert_{S,D}$ & $J/\psi\,\omega\vert_P$       & $-$   
        &  $\chi_{cJ}\,\pi \vert_{S,D}$ &$J/\psi\,\rho\vert_P$  \\
        & $-$                   & 3655                  &3874                           & $-$   
        & 3655                          & 3874\\
\hline
$2^{--}$&$-$                    & $J/\psi\,\eta\vert_P$         & $D^*\,\bar D^*\vert_P$                & $-$   
        & $J/\psi\,\pi\vert_P$          &$D^*\,\bar D^*\vert_P$ \\
        & $-$                   & 3233                  & 4040                          & $-$   
        & 3233                          & 4040          \\
\hline
$3^{-+}$& $-$                   &$-$                    & $J/\psi\,\omega\vert_P$       & $-$   
        & $-$                           &$J/\psi\,\rho\vert_P$  \\
        & $-$                   & $-$                   & 3874                          & $-$   
        & $-$                           & 3874\\
\hline
$3^{--}$& $-$                   & $-$                   &$D^*\,\bar D^*\vert_P$         & $-$   
        & $-$                           &$D^*\,\bar D^*\vert_P$ \\
        & $-$                   & $-$                   & 4040                          & $-$   
        & $-$                           & 4040          \\
\hline
\end{tabular}
\end{table}
\end{center}

\begin{center}
\begin{table}[htb]
\caption{$(L,S)=(0,1)$ $J^{PC}=1^{++}$ lowest two-meson thresholds 
for $S-$, $P-$, and $D-$wave final state relative angular momentum. 
Both possible couplings, $(c\bar n)(n\bar c)$ and $(c \bar c)(n \bar n)$,
are considered. Energies are in MeV.}
\label{th5}
\begin{tabular}{|c|ccc|ccc|}
\hline
                & \multicolumn{3}{c|}{$(c\bar n)(n\bar c)$}     & \multicolumn{3}{c|}{$(c\bar c)(n\bar n)$}\\
\hline
Experiment      & $D\,\bar D^*\vert_S$  & $D\,\bar D_0^*\vert_P$& $D_1\,\bar D_0^*\vert_D$      &
                $J/\psi\,\omega\vert_S$ & $\eta_c\, f_0\vert_P$ & $\chi_{c1}\,f_0 \vert_D$ \\
                & 3871                  & 4176                  & 4731                          &
                3880                    & 3580                  & 4111 \\
\hline
CQC             & $D\,\bar D^*\vert_S$  &$D\,\bar D_J^*\vert_P$ &$D_1\,\bar D_J^*\vert_D$&
                $J/\psi\,\omega\vert_S$ &$h_c\,\omega\vert_P$   &$\chi_{cJ}\,f_J\vert_D$        \\
                &3937                   &4434                   &4988                   &
                3745                    & 4157  & 4771 \\
\hline
BCN             & $D\,\bar D^*\vert_S$ & $D\,\bar D_J^*\vert_P$ & $D_1\,\bar D_J^*\vert_D$
                &  $J/\psi\,\omega\vert_S$ &  $\chi_{cJ}\,\eta \vert_P$&$\chi_{cJ}\,f_J \vert_D$\\ 
                & 3906 & 4377 &  4946 &  3874& 3655& 4773\\
\hline
\end{tabular}
\end{table}
\end{center}

\begin{center}
\begin{table}[htb]
\caption{Energy, $E_{4q}\equiv E_{4q} (K_{\rm max})$ (in MeV) and probability, $P_T$, of the
basis vector corresponding to the lowest threshold for CQC and BCN models. $T(M_1,M_2)$ indicates
the lowest physical threshold, $E_T$ its energy as obtained from Tables \protect\ref{th2},
\protect\ref{th3}, and \protect\ref{th4} and $\Delta_E$ is defined in Eq. (\protect\ref{delta}).
We also quote in the last three columns the experimental thresholds.}
\label{tC1}
\begin{tabular}{|cc|cc|ccc|ccc|}
\hline
$K_{\rm max}$ & $(L,S)\,J^{PC}$ & $E_{4q}$      & $P_T$  &   T$(M_1,M_2)$               & $E_T$ & $\Delta_E$
&   T$(M_1,M_2)$        & $E_T$ & $\Delta_E$\\
\hline
\multicolumn{2}{|c|}{$\,$}& \multicolumn{2}{c|}{CQC}&
\multicolumn{3}{c|}{CQC}&\multicolumn{3}{c|}{Exp.}\\
\hline
24        & $(0,0)\,0^{++}$&  3779      & 0.9954 & $J/\psi\,\omega\vert_S$      & 3745  & +34  & 
$\eta_c\,\eta\vert_S$   & 3528  &   +251 \\
22        &$(0,0)\,0^{+-}$ &  4224      & 0.9995 & $\chi_{cJ}\,\omega\vert_P$   & 4160  & +64 &
$J/\psi f_0\vert_P$     & 3697  & +438 \\
20        & $(0,1)\,1^{++}$&  3786      & 0.9968 & $J/\psi\,\omega\vert_S$      & 3745  & +41 &
$\eta_c\,f_0\vert_P$    & 3580  & +206 \\
22        &$(0,1)\,1^{+-}$ &  3728      & 0.9983 & $\eta_c\,\omega\vert_P$      & 3683  & +45 &
$J/\psi\,\eta\vert_S$   & 3644 & +84\\
28        & $(0,2)\,2^{++}$&  3774      & 0.9989 & $J/\psi\,\omega\vert_S$      & 3745  & +29 &
$J/\psi\,\omega\vert_S$ & 3880 & $-$106\\
28        &$(0,2)\,2^{+-}$ &  4214      & 0.9997 & $\chi_{cJ}\,\omega\vert_P$   & 4160  & +54 &
$J/\psi\,f_0\vert_P$    & 3697 & +517\\
19        &$(1,0)\,1^{-+}$ &  3829      & 0.9998 & $J/\psi\,\omega\vert_P$      & 3745  & +84 &
$\eta_c\,\eta\vert_P$   & 3528 & +301\\
19        &$(1,0)\,1^{--}$ &  3969      & 0.9451 & $D\,\bar D\vert_P$           & 3872  & +97 & 
$J/\psi\,f_0\vert_{S,D}$& 3697 & +272\\
17        &$(1,1)\,(0,1,2)^{-+}$ & 3839 & 0.9998 & $J/\psi\,\omega\vert_P$      & 3745  & +94 &
$\eta_c\,f_0\vert_{S,D}$\footnote{$S-$wave for the $J=0$ threshold and $D-$wave for the $J=2$} & 3580 & +259\\
&&&&&&&$D\,\bar D^*\vert_P$\footnote{$J=1$ state} & 3871 & $-$32\\
17        &$(1,1)\,(0,1,2)^{--}$ & 3791 & 0.9997 & $\eta_c\,\omega\vert_P$      & 3683  & +108 &
$J/\psi\,\eta\vert_P$   & 3644  & +147\\
21        &$(1,2)\,(1,2,3)^{-+}$ & 3820 & 0.9999 & $J/\psi\,\omega\vert_S$      & 3745  & +75 &
$J/\psi\,\omega\vert_P$ & 3880  & $-$60\\
21        &$(1,2)\,(1,2,3)^{--}$ & 4054 & 0.9999 & $D^*\,\bar D^*\vert_P$       & 4002  & +52 &
$J/\psi\,f_0\vert_D$    & 3697  &+357 \\
\hline
\hline
\multicolumn{2}{|c|}{$\,$}& \multicolumn{2}{c|}{BCN}&
\multicolumn{3}{c|}{BCN}&\multicolumn{3}{c|}{Exp.}\\
\hline
$K_{\rm max}$ & $(L,S)\,J^{PC}$ & $E_{4q}$      & $P_T$  &   T$(M_1,M_2)$               & $E_T$ & $\Delta_E$
&   T$(M_1,M_2)$        & $E_T$ & $\Delta_E$\\
\hline
26        & $(0,0)\,0^{++}$ &  3249     & 0.9993 &  $\eta_c\,\eta\vert_S$       & 3174  & +75 &
$\eta_c\,\eta\vert_S$   & 3528  &   $-$279 \\
24        &$(0,0)\,0^{+-}$  &  3778     & 0.9997 &  $h_c\,\eta\vert_P$          & 3638  & +140 &
$J/\psi f_0\vert_P$     & 3697  & +81 \\
22        & $(0,1)\,1^{++}$ &  3808     & 0.9997 &  $\chi_{cJ}\,\eta \vert_P$   & 3655  & +153 &
$\eta_c\,f_0\vert_P$    & 3580  & +228 \\
22        &$(0,1)\,1^{+-}$  &  3319     & 0.9993 &  $J/\psi\,\eta\vert_S$       & 3233  & +86 &
$J/\psi\,\eta\vert_S$   & 3644 & $-$325\\
26        & $(0,2)\,2^{++}$ &  3897     & 0.9987 &  $J/\psi\,\omega\vert_S$     & 3874  & +23 &
$J/\psi\,\omega\vert_S$ & 3880 & +17\\
28        &$(0,2)\,2^{+-}$  &  4328     & 0.9998 &  $\chi_{cJ}\,\omega\vert_P$  & 4296  & +32 &
$J/\psi\,f_0\vert_P$    & 3697 & +631\\
21        &$(1,0)\,1^{-+}$  &  3331     & 0.9999 &  $\eta_c\,\eta\vert_P$       & 3174  & +157 &
$\eta_c\,\eta\vert_P$   & 3528 & $-$197\\
21        &$(1,0)\,1^{--}$  &  3732     & 0.9934 &  $h_c\,\eta\vert_{S,D}$      & 3638  & +94 &
$J/\psi\,f_0\vert_{S,D}$& 3697 & +35\\
19        &$(1,1)\,(0,1,2)^{-+}$ & 3760 & 0.9950 &  $\chi_{cJ}\,\eta \vert_{S,D}$ & 3655& +105 &
$\eta_c\,f_0\vert_{S,D^1}$ & 3580 & +180 \\
&&&&&&&$D\,\bar D^*\vert_{P^2}$ & 3871 & $-$111\\
19        &$(1,1)\,(0,1,2)^{--}$ & 3405 & 0.9998 &  $J/\psi\,\eta\vert_P$       & 3233  & +172 &
$J/\psi\,\eta\vert_P$   & 3644  & $-$239\\
21         &$(1,2)\,(1,2,3)^{-+}$ & 3929& 0.9999 &  $J/\psi\,\omega\vert_S$     & 3874  & +55 &
$J/\psi\,\omega\vert_P$ & 3880  & +49\\
21        &$(1,2)\,(1,2,3)^{--}$ & 4092 & 0.9999 &  $D^*\,\bar D^*\vert_P$      & 4040  & +52 & 
$J/\psi\,f_0\vert_D$    & 3697  &+395 \\
\hline
\end{tabular}
\end{table}
\end{center}

\begin{center}
\begin{table}[htb]
\caption{Energy (MeV), radius (fm), and probability of the different components
of the four-quark wave function as a function of $K$ for the 
$J^{PC}=2^{++}$ and $J^{PC}=2^{+-}$ $(L,S)=(0,2)$ channels. The lowest 
threshold is quoted in the bottom part of the table.}
\label{tC3}
\begin{tabular}{|c|cc|cc||c|cc|cc|}
\hline
&\multicolumn{4}{c||}{$J^{PC}=2^{++}$}&&\multicolumn{4}{c|}{$J^{PC}=2^{+-}$}\\
\hline
$K$     & E & RMS & $P_{11}(1,1)$       & $P_{88}(1,1)$ &
$K$     & E & RMS & $P_{11}(1,1)$       & $P_{88}(1,1)$ \\
\hline
\multicolumn{10}{|c|}{CQC}\\
\hline
0       & 4140  & 0.3418        & 1.0000        & 0.0000 &
0       & $-$   & $-$   & $-$   & $-$ \\
2       & 3986  & 0.3697        & 0.9890        & 0.0110 &
2       & 4689  & 0.4851        & 0.9998        & 0.0002        \\
4       & 3912  & 0.4010        & 0.9884        & 0.0116 &
4       & 4508  & 0.5183        & 0.9962        & 0.0038        \\
6       & 3872  & 0.4337        & 0.9913        & 0.0087        &
6       & 4414  & 0.5567        & 0.9961        & 0.0039        \\
8       & 3846  & 0.4669        & 0.9934        & 0.0066        &
8       & 4359  & 0.5967        & 0.9974        & 0.0026        \\
10      & 3828  & 0.5001        & 0.9951        & 0.0049        &
10      & 4321  & 0.6372        & 0.9981        & 0.0019        \\
12      & 3815  & 0.5329        & 0.9962        & 0.0038        &
12      & 4295  & 0.6775        & 0.9987        & 0.0013        \\
14      & 3805  & 0.5652        & 0.9970        & 0.0030        &
14      & 4275  & 0.7175        & 0.9990        & 0.0010        \\
16      & 3798  & 0.5970        & 0.9975        & 0.0025        &
16      & 4260  & 0.7570        & 0.9992        & 0.0008        \\
18      & 3792  & 0.6283        & 0.9980        & 0.0020        &
18      & 4248  & 0.7959        & 0.9994        & 0.0006        \\
20      & 3787  & 0.6590        & 0.9983        & 0.0017        &
20      & 4239  & 0.8342        & 0.9995        & 0.0005        \\
22      & 3783  & 0.6889        & 0.9985        & 0.0015        &
22      & 4231  & 0.8718        & 0.9996        & 0.0004    \\
24      & 3779  & 0.7185        & 0.9987        & 0.0013        &
24      & 4224  & 0.9087        & 0.9997        & 0.0003    \\
26      & 3776  & 0.7475        & 0.9989        & 0.0011        &
26      & 4219  & 0.9451        & 0.9997        & 0.0003    \\
28      & 3774  & $-$           & $-$           & $-$           &
28      & 4214  & $-$           & $-$           & $-$           \\
\hline
\hline
$J/\psi\,\omega\vert_S$ & 3745 & 0.5745 & 1             & 0 &
$\chi_{cJ}\,\omega \vert_P$ & 4160&0.6873       & 1             & 0 \\
\hline
\hline
\multicolumn{10}{|c|}{BCN}\\
\hline
0       & 4196          & 0.3393        & 1.0000        & 0.0000        & &&&& \\
2       & 4057          & 0.3778        & 0.9862        & 0.0138        &
2       & 4732          & 0.4628        & 0.9943        & 0.0057        \\
4       & 3999          & 0.4168        & 0.9865        & 0.0135        &
4       & 4557          & 0.5127        & 0.9929        & 0.0071        \\
6       & 3968          & 0.4568        & 0.9899        & 0.0101        &
6       & 4478          & 0.5639        & 0.9939        & 0.0061        \\
8       & 3948          & 0.4972        & 0.9923        & 0.0077        &
8       & 4433          & 0.6163        & 0.9964        & 0.0036        \\
10      & 3935          & 0.5375        & 0.9944        & 0.0056        &
10      & 4405          & 0.6694        & 0.9977        & 0.0023        \\
12      & 3925          & 0.5775        & 0.9957        & 0.0043        &
12      & 4385          & 0.7226        & 0.9986        & 0.0014        \\
14      & 3918          & 0.6168        & 0.9966        & 0.0034        &
14      & 4371          & 0.7758        & 0.9990        & 0.0010        \\
16      & 3913          & 0.6561        & 0.9973        & 0.0027        &
16      & 4360          & 0.8288        & 0.9993        & 0.0007        \\
18      & 3908          & 0.6946        & 0.9978        & 0.0022        &
18      & 4352          & 0.8816        & 0.9995        & 0.0005        \\
20      & 3905          & 0.7309        & 0.9981        & 0.0019        &
20      & 4345          & 0.9341        & 0.9996        & 0.0004        \\
22      & 3902          & 0.7720        & 0.9984        & 0.0016        &
22      & 4340          & 0.9863        & 0.9997        & 0.0003        \\
24      & 3899          & 0.8091        & 0.9987        & 0.0013        &
24      & 4335          & 1.0382        & 0.9998        & 0.0002        \\
26      & 3897          & $-$           & $-$           & $-$           &
26      & 4332          & 1.0899        & 0.9998        & 0.0002        \\
28      &               &               &               &               &
28      & 4328          & $-$           & $-$           & $-$           \\
\hline
\hline
$J/\psi\,\omega\vert_S$ & 3874 &0.6133 &  1 & 0 &
$\chi_{cJ}\,\omega \vert_P$ & 4296 & 0.7259 & 1 & 0 \\
\hline
\end{tabular}
\end{table}
\end{center}

\begin{center}
\begin{table}[htb]
\caption{Energies (in MeV) obtained using the extrapolation Eq. (\protect\ref{extra})
for the states of Table  \protect\ref{tC3}.}
\label{tC4}
\begin{tabular}{|cc|cc|}
\hline
\multicolumn{2}{|c|}{$J^{PC}=2^{++}$} & \multicolumn{2}{c|}{$J^{PC}=2^{+-}$} \\
\hline
$(K_o,K_f)$     & $E(K=\infty)$& $(K_o,K_f)$    & $E(K=\infty)$ \\
\hline
\multicolumn{4}{|c|}{CQC}\\
\hline
(2,28)          &       3657    & (2,28)        & 4042 \\
(4,28)          &       3704    & (4,28)        & 4106 \\
(6,28)          &       3720    & (6,28)        & 4124 \\
(8,28)          &       3727    & (8,28)        & 4135 \\
(10,28)         &       3731    & (10,28)       & 4141 \\
(12,28)         &       3734    & (12,28)       & 4145 \\
(14,28)         &       3736    & (14,28)       & 4148 \\
(16,28)         &       3737    & (16,28)       & 4150 \\
(18,28)         &       3739    & (18,28)       & 4151 \\
(20,28)         &       3739    & (20,28)       & 4153 \\
(22,28)         &       3740    & (22,28)       & 4154 \\
(24,28)         &       3741    & (24,28)       & 4156 \\
\hline
Threshold       & 3745          & Threshold     & 4160 \\
\hline
\hline
\multicolumn{4}{|c|}{BCN}\\
\hline
(2,26)          & 3816          & (2,28)        &  4242\\
(4,26)          & 3848          & (4,28)        &  4268\\
(6,26)          & 3859          & (6,28)        &  4277\\
(8,26)          & 3864          & (8,28)        &  4282\\
(10,26)         & 3867          & (10,28)       &  4286\\
(12,26)         & 3869          & (12,28)       &  4288\\
(14,26)         & 3871          & (14,28)       &  4290\\
(16,26)         & 3872          & (16,28)       &  4291\\
(18,26)         & 3872          & (18,28)       &  4292\\
(20,26)         & 3873          & (20,28)       &  4293\\
(22,26)         & 3874          & (22,28)       &  4294\\
                &               & (24,28)       &  4296\\
\hline
Threshold       & 3874                          & Threshold     & 4296 \\
\hline
\end{tabular}
\end{table}
\end{center}

\begin{center}
\begin{table}[htb]
\caption{Comparison of the energies (MeV) and probability of the dominant components
of the four-quark wave function for $J^{PC}=1^{++}$ $(L,S)=(0,1)$ ($K=8$) and 
$J^{PC}=(1,2,3)^{--}$ $(L,S)=(1,2)$ ($K=7$) states either using $\sum_i \ell_i=\infty$ 
or $\sum_i \ell_i\leq1$ using the CQC model.}
\label{tlmax}
\begin{tabular}{|c||cc|cc|}
\hline
                & \multicolumn{2}{c|}{$\sum_i \ell_i=\infty$} & \multicolumn{2}{c|}{$\sum_i \ell_i\leq1$} \\
\cline{2-5}
                & Energy        & Probability           & Energy        & Probability           \\
\hline
$L=0$ $S=1$     & 3844          & $P_{11}(1,1)=0.9871$  & 3850          & $P_{11}(1,1)=1.0000$  \\
$L=1$ $S=2$     & 4199          & $P_{88}(1,1)=0.8779$  & 4275          & $P_{11}(1,1)=0.7088$  \\
\hline
\end{tabular}
\end{table}
\end{center}

\begin{center}
\begin{table}[htb]
\caption{Energy (MeV), radius (fm) and probability of the different components
of the four-quark wave function as a function of $K$ for the 
$J^{PC}=1^{--}$ $(L,S)=(1,0)$ using the CQC model.}
\label{taux1}
\begin{tabular}{|c|cc|cccc|}
\hline
$K$     & E & RMS & $P_{11}(0,0)$       & $P_{11}(1,1)$ & $P_{8,8}(0,0)$        & $P_{8,8}(1,1)$ \\
\hline
1       & 4432  & 0.3939        & 0.0068        & 0.1342        & 0.0068        & 0.8522 \\
3       & 4228  & 0.4242        & 0.0037        & 0.0974        & 0.0108        & 0.8882 \\
5       & 4132  & 0.4600        & 0.0039        & 0.1032        & 0.0175        & 0.8754 \\
7       & 4078  & 0.4969        & 0.0046        & 0.1031        & 0.0264        & 0.8660  \\
9       & 4044  & 0.5341        & 0.0056        & 0.1029        & 0.0363        & 0.8552  \\
11      & 4020  & 0.5714        & 0.0066        & 0.1018        & 0.0464        & 0.8452  \\
13      & 4002  & 0.6086        & 0.0076        & 0.1009        & 0.0556        & 0.8358  \\
15      & 3989  & 0.6458        & 0.0085        & 0.1001        & 0.0637        & 0.8276  \\
17      & 3978  & 0.6828        & 0.0093        & 0.0996        & 0.0708        & 0.8203  \\
\hline
\hline
$D\,\bar D\vert_P$  & 3872& 0.4396      & 0.0278        & 0.0833        & 0.2222        & 0.6667 \\
\hline
\end{tabular}
\end{table}
\end{center}

\begin{center}
\begin{table}[htb]
\caption{Energy (MeV), radius (fm) and probability of the different
components of the four-quark wave function as a function of $K$ for the 
$J^{PC}=(1,2,3)^{--}$ $(L,S)=(1,2)$ using the CQC and BCN models.}
\label{taux2}
\begin{tabular}{|c|cc|cc||c|cc|cc|}
\hline
&\multicolumn{4}{c||}{CQC}&&\multicolumn{4}{c|}{BCN}\\
\hline
$K$     & E & RMS & $P_1(1,1)$  & $P_8(1,1)$    &
$K$     & E & RMS & $P_1(1,1)$  & $P_8(1,1)$    \\
\hline
1       & 4476  & 0.4063        & 0.2116        & 0.7884        &
1       & 4518  & 0.3965        & 0.2348        & 0.7652 \\
3       & 4287  & 0.4398        & 0.1488        & 0.8512 &
3       & 4332  & 0.4442        & 0.1671        & 0.8329 \\
5       & 4199  & 0.4790        & 0.1438        & 0.8562 &
5       & 4247  & 0.4919        & 0.1374        & 0.8626 \\
7       & 4151  & 0.5193        & 0.1333        & 0.8666 &
7       & 4199  & 0.5412        & 0.1221        & 0.8779 \\
9       & 4121  & 0.5601        & 0.1270        & 0.8729 &
9       & 4167  & 0.5909        & 0.1163        & 0.8837 \\
11      & 4101  & 0.6011        & 0.1218        & 0.8782 &
11      & 4145  & 0.6406        & 0.1134        & 0.8866 \\
13      & 4086  & 0.6424        & 0.1185        & 0.8815 &
13      & 4128  & 0.6900        & 0.1121        & 0.8879 \\
15      & 4075  & 0.6839        & 0.1162        & 0.8838 &
15      & 4116  & 0.7389        & 0.1114        & 0.8886 \\
17      & 4066  & 0.7257        & 0.1147        & 0.8853 &
17      & 4106  & 0.7872        & 0.1111        & 0.8889 \\
19      & 4059  & 0.7675        & 0.1136        & 0.8864 &
19      & 4099  & 0.8351        & 0.1111        & 0.8889 \\
21      & 4054  & 0.8094        & 0.1129        & 0.8871 &
21      & 4092  & 0.8824        & 0.1111        & 0.8889 \\
\hline
\hline
$D^*\,\bar D^*\vert_P$          & 4002 & 0.4684 & 0.1111        & 0.8889 &
$D^*\,\bar D^*\vert_P$     & 4040&0.4794        & 0.1111        & 0.8889 \\
$\chi_{cJ}\,\omega \vert_P$ & 4160 & 0.6873 & 1 & $-$ &
$\chi_{cJ}\,\omega \vert_{S,D}$ & 4296&0.7259   & 1             & $-$ \\
\hline
\end{tabular}
\end{table}
\end{center}

\begin{center}
\begin{table}[htb]
\caption{Energy (MeV), radius (fm) and probability of the dominant component
of the four-quark wave function as a function of $K$ 
for two different parametrizations of the
CQC $(L,S)=(0,0)$ $J^{PC}=0^{++}$ and BCN $(L,S)=(1,0)$ $J^{PC}=1^{--}$.}
\label{tevol}
\begin{tabular}{|c|cc|cc||c|cc|cc|}
\hline
\multicolumn{10}{|c|}{CQC $J^{PC}=0^{++}$ $(L,S)=(0,0)$}\\
\hline
&\multicolumn{4}{c||}{$r^{n\bar n}_0=0.38$}&&\multicolumn{4}{c|}{$r^{n\bar n}_0=0.18$}\\
\hline
$K$     &E & RMS        & $P_{11}(1,1)$ &  $P_{11}(0,0)$ &
$K$     &E & RMS        & $P_{11}(1,1)$ &  $P_{11}(0,0)$\\
\hline
2       & 3984  & 0.3685        & 0.9685        &  0.0024 &
2       & 3963  & 0.3649        & 0.0011        &  0.9838 \\
4       & 3909  & 0.3987        & 0.9625        & 0.0014 &
4       & 3889  & 0.3957        & 0.0010        & 0.9853 \\
6       & 3849  & 0.4282        & 0.9898        &  0.0007 &
6       & 4414  & 0.5562        & 0.0005        & 0.9948 \\
8       & 3843  & 0.4629        & 0.9766        &  0.0004 &
8       & 3822  & 0.4608        & 0.0003        & 0.9923 \\
10      & 3826  & 0.4955        & 0.9827        &  0.0002 &
10      & 3804  & 0.4932        & 0.0002        & 0.9941  \\
12      & 3814  & 0.5280        & 0.9868        &  0.0001 &
12      & 3791  & 0.5251        & 0.0001        & 0.9957 \\
14      & 3804  & 0.5602        & 0.9898        &  0.0001 &
14      & 3782  & 0.5564        & 0.0001        & 0.9966 \\
\hline
\hline
$J/\psi\,\omega\vert_S$         & 3745& 0.5745 &  1     &  0 &
$\eta_c\,\eta\vert_S$           & 3718& 0.5749 &  0     &  1 \\
\hline
\hline
\multicolumn{10}{|c|}{BCN $J^{PC}=1^{--}$ $(L,S)=(1,0)$}\\
\hline
&\multicolumn{4}{c||}{$r_0=2.2$ fm}&
&\multicolumn{4}{c|}{$r_0=0.5$ fm}\\
\hline
$K$     &E & RMS        & $P_{11}$      & $P_{88}$ &
$K$     &E & RMS        & $P_{11}$      & $P_{88}$ \\
\hline
1       & 4290  & 0.3563        & 0.9267 & 0.0733 &
1       & 4467  & 0.3899        & 0.5344 & 0.4656 \\
3       & 4066  & 0.3793        & 0.9363 & 0.0637 &
3       & 4276  & 0.4341        & 0.2734 & 0.7266 \\
5       & 3954  & 0.3993        & 0.9559 & 0.0441 &
5       & 4188  & 0.4791        & 0.1817 & 0.8182 \\
7       & 3888  & 0.4185        & 0.9702 & 0.0298 &
7       & 4137  & 0.5261        & 0.1449 & 0.8551 \\
9       & 3843  & 0.4371        & 0.9791 & 0.0209 &
9       & 4103  & 0.5735        & 0.1298 & 0.8702 \\
11      & 3811  & 0.4553        & 0.9847 & 0.0153 &
11      & 4080  & 0.6207        & 0.1220 & 0.8780 \\
13      & 3787  & 0.4731        & 0.9883 & 0.0117 &
13      & 4062  & 0.6676        & 0.1180 & 0.8820 \\
\hline
\hline
$h_c\,\eta\vert_P$              & 3638& 0.5764 &  1     & 0 &
$D\,\bar D\vert_P$              & 3961& 0.4567 &  0.1111 & 0.8889\\
\hline
\end{tabular}
\end{table}
\end{center}

\begin{center}
\begin{table}[htb]
\caption{Energy (MeV), radius (fm) and probability of the dominant 
components of the four-quark wave function as a function of $K$ for
$J^{PC}=1^{++}$ $(L,S)=(0,1)$ and 
$J^{PC}=2^{-+}$ $(L,S)=(1,1)$ states both for CQC and BCN models.}
\label{tX1}
\begin{tabular}{|c|cc|cc||c|cc|cc|}
\hline
\multicolumn{10}{|c|}{$J^{PC}=1^{++}$ $(L,S)=(0,1)$}\\
\hline
&\multicolumn{4}{c||}{CQC}&&\multicolumn{4}{c|}{BCN}\\
\hline
$K$     & E     & RMS   & $P_{11}(1,1)$ & $P_{88}(1,1)$ &
$K$     & E     & RMS   & $P_{11}(1,0)$ & $P_{11}(1,1)$ \\
\hline
0       & 4141  & 0.3418        & 1.0000        & 0.0000 &
0       & 4196  & 0.3393        & 0.0000        & 1.0000        \\
2       & 3985  & 0.3692        & 0.9822        & 0.0178 &
2       & 4053  & 0.3766        & 0.0000        & 0.9462        \\
4       & 3911  & 0.4000        & 0.9789        & 0.0211 &
4       & 3994  & 0.4133        & 0.0000        & 0.9233        \\
6       & 3870  & 0.4322        & 0.9834        & 0.0166        &
6       & 3963  & 0.4502        & 0.0000        & 0.9236        \\
8       & 3845  & 0.4650        & 0.9871        & 0.0129        &
8       & 3944  & 0.4883        & 0.0001        & 0.9302        \\
10      & 3827  & 0.4979        & 0.9905        & 0.0095        &
10      & 3932  & 0.5267        & 0.0002        & 0.9424        \\
12      & 3814  & 0.5305        & 0.9926        & 0.0074        &
12      & 3920  & 0.5581        & 0.9321        & 0.0605        \\
14      & 3805  & 0.5628        & 0.9943        & 0.0057        &
14      & 3887  & 0.5829        & 0.9986        & 0.0004        \\
16      & 3797  & 0.5945        & 0.9954        & 0.0046        &
16      & 3861  & 0.6063        & 0.9993        & 0.0001        \\
18      & 3791  & 0.6255        & 0.9962        & 0.0038        &
18      & 3840  & 0.6298        & 0.9995        & 0.0000        \\
20      & 3786  & 0.6564        & 0.9968        & 0.0032        &
20      & 3822  & 0.6520        & 0.9996        & 0.0000        \\
22      & $-$   & $-$   & $-$   & $-$   &
22      & 3808  & 0.6736        & 0.9997        & 0.0000        \\
\hline
\hline
$J/\psi\,\omega\vert_S$ & 3745  & 0.5745 & 1    & 0 &
$\chi_{cJ}\,\eta \vert_P$& 3655 & 0.5814 & 1    & 0             \\
$h_c\,\omega\vert_P$    & 4157  & 0.6857 & 0    & 0 &
$J/\psi\,\omega\vert_S$  & 3874 & 0.6133 & 0    & 1             \\
$\chi_{cJ}\,f_J\vert_D$ & 4771  & 0.9706 & 1    & 0 &
$\chi_{cJ}\,f_J \vert_D$ & 4773 & 0.8833 & 0    & 1             \\
\hline
\hline
\multicolumn{10}{|c|}{$J^{PC}=2^{-+}$ $(L,S)=(1,1)$}\\
\hline
&\multicolumn{4}{c||}{CQC}&&\multicolumn{4}{c|}{BCN}\\
\hline
$K$     & E & RMS       & $P_{11}(1,1)$ & $P_{88}(1,1)$ &
$K$     & E & RMS       & $P_{11}(1,0)$ & $P_{88}(1,0)$ \\
\hline
1       & 4311  & 0.4172        & 1.0000        & 0.0000 &
1       & 4315  & 0.3599        & 0.9627        & 0.0302        \\
3       & 4117  & 0.4521        & 0.9982        & 0.0017 &
3       & 4088  & 0.3828        & 0.9615        & 0.0343        \\
5       & 4018  & 0.4926        & 0.9986        & 0.0014 &
5       & 3975  & 0.4029        & 0.9715        & 0.0260        \\
7       & 3958  & 0.5347        & 0.9991        & 0.0009 &
7       & 3908  & 0.4222        & 0.9800        & 0.0184        \\
9       & 3918  & 0.5768        & 0.9994        & 0.0006 &
9       & 3862  & 0.4409        & 0.9856        & 0.0134        \\
11      & 3889  & 0.6186        & 0.9996        & 0.0004 &
11      & 3830  & 0.4591        & 0.9892        & 0.0101        \\
13      & 3868  & 0.6596        & 0.9997        & 0.0003 &
13      & 3806  & 0.4767        & 0.9915        & 0.0079        \\
15      & 3852  & 0.6999        & 0.9998        & 0.0002 &
15      & 3787  & 0.4939        & 0.9931        & 0.0065        \\
17      & 3839  & $-$           & $-$           & $-$    &
17      & 3772  & 0.5106        & 0.9942        & 0.0054        \\
19      & $-$   & $-$           & $-$           & $-$    &
19      & 3760  & 0.5270        & 0.9950        & 0.0047        \\
\hline
\hline
$J/\psi\,\omega\vert_P$ & 3745& 0.5745  & 1             & 0 &
$\chi_{cJ}\,\eta \vert_{S,D}$   & 3655 & 0.5814 & 1 & 0 \\
$h_c\,\omega\vert_{S,D}$& 4157& 0.6857  & 0             & 0 &
$J/\psi\,\omega\vert_P$         & 3874 & 0.6135 & 0 & 0 \\
\hline
\end{tabular}
\end{table}
\end{center}



\begin{thebibliography}{99}

\bibitem{Bel03} Belle Collaboration, S.-K. Choi {\it et al.},
                Phys. Rev. Lett. {\bf 91}, 262001 (2003).

\bibitem{Bab05b} BaBar Collaboration, B. Aubert {\it et al.},
                Phys. Rev. D {\bf 71}, 071103R (2005).

\bibitem{CDF04} CDF Collaboration, D. Acosta {\it et al.},
                Phys. Rev. Lett. {\bf 93}, 072001 (2004).

\bibitem{D0c05} D0 Collaboration, V.M. Abazov {\it et al.},
                Phys. Rev. Lett. {\bf 93}, 162002 (2004).

\bibitem{Bel4b} Belle Collaboration, S.-K. Choi {\it et al.},
                Phys. Rev. Lett. {\bf 94}, 182002 (2005).

\bibitem{Bel05} Belle Collaboration, K. Abe {\it et al.},
                Phys. Rev. Lett. {\bf 98}, 082001 (2007).

\bibitem{Bel5b} Belle Collaboration, S. Uehara {\it et al.},
                Phys. Rev. Lett. {\bf 96}, 082003 (2006).

\bibitem{Bab05} BaBar Collaboration, B. Aubert {\it et al.},
                Phys. Rev. Lett. {\bf 95}, 142001 (2005).

\bibitem{Gel80} G. Gelmini,
                Nucl. Phys. B {\bf 174}, 509 (1980).

\bibitem{Cha80} K.T. Chao,
                Nucl. Phys. B {\bf 169}, 281 (1980).

\bibitem{Cha77} H.M. Chan and H. Hogaasen,
                Phys. Lett. B {\bf 72}, 121 (1977).

\bibitem{Cha81} K.T. Chao,
                Nucl. Phys. B {\bf 183}, 435 (1981).

\bibitem{Sil93} B. Silvestre-Brac and C. Semay, 
                Z. Phys. C {\bf 57}, 273 (1993).

\bibitem{Mai05} L. Maiani, F. Piccinini, A.D. Polosa, and V. Riquer,
                Phys. Rev. D {\bf 71}, 014028 (2005).

\bibitem{Ebe05} D. Ebert, R.N. Faustov, and O. Galkin,
                Phys. Lett. B {\bf 634}, 214 (2006).

\bibitem{Hog06} H. Hogaasen, J.-M. Richard, and P. Sorba,
                Phys. Rev. D {\bf 73}, 054013 (2006).

\bibitem{Mat07} R.D. Matheus, S. Narison, M. Nielsen, and J.-M. Richard,
                Phys. Rev. D {\bf 75}, 014005 (2007).

\bibitem{Bab03} BaBar Collaboration, B. Aubert {\it et al.},
                        Phys. Rev. Lett. {\bf 90}, 242001 (2003).

\bibitem{Cle03} CLEO Collaboration, D. Besson {\it et al.},
                        Phys. Rev. D {\bf 68}, 032002 (2003).

\bibitem{Bel04} Belle Collaboration, Y. Mikani {\it et al.},
                        Phys. Rev. Lett. {\bf 92}, 012002 (2004).

\bibitem{Belb4} Belle Collaboration, K. Abe {\it et al.}, 
                Phys. Rev. D {\bf 69}, 112002 (2004).

\bibitem{Foc04} FOCUS Collaboration, J.M. Link {\it et al.},
                        Phys. Lett. B {\bf 586}, 11 (2004).

\bibitem{Swa06} E.S. Swanson,
                Phys. Rep. {\bf 429}, 243 (2006) and references therein.

\bibitem{Oka06} M. Oka, 
                Nucl. Phys. A {\bf 790}, 462c (2007).

\bibitem{Vij06}J. Vijande, F. Fern\'andez, and A. Valcarce,
                Phys. Rev. D {\bf 73}, 034002 (2006).

\bibitem{Jaf07} R.L. Jaffe, 
                hep-ph/0701038.

\bibitem{Vij05b} J. Vijande, A. Valcarce, F. Fern\'andez, and B.  Silvestre-Brac, 
                Phys. Rev. D {\bf 72}, 034025 (2005).

\bibitem{Bar06} N. Barnea, J. Vijande, and A. Valcarce,
                Phys. Rev. D {\bf 73}, 054004 (2006).

\bibitem{Nir9798} N. Barnea and A. Novoselsky,
                        Ann. Phys. (N. Y.) {\bf 256}, 192 (1997);
                        Phys. Rev. A {\bf 57}, 48 (1998).

\bibitem{Jaf77} R.L. Jaffe,
                Phys. Rev. D {\bf 15}, 267 (1977); 281 (1977).

\bibitem{Fab83} M. Fabre de la Ripelle,
                Ann. Phys. (N. Y.) {\bf 147}, 281 (1983).

\bibitem{Efr72} V.D. Efros, 
                Yad. Fiz. {\bf 15}, 226 (1972) 
                [Sov. J. Nucl. Phys. {\bf 15}, 128 (1972)].

\bibitem{Efr95} V. D. Efros,
                Few-Body Systems {\bf 19}, 169 (1995). 

\bibitem{PDG06} W.-M. Yao {\it et al.}, 
                J. Phys. G {\bf 33} 1 (2006).

\bibitem{Bha81} R.K. Bhaduri, L.E. Cohler, and Y. Nogami,
                Nuovo Cimento {\bf A65}, 376 (1981)

\bibitem{Sil85} B. Silvestre-Brac and C. Gignoux,
                Phys. Rev. D {\bf 32}, 743 (1985).

\bibitem{Ruj75} A. de R\'ujula, H. Georgi, and S.L. Glashow,
                Phys. Rev. D {\bf 12}, 147 (1975).

\bibitem{Rep05} A. Valcarce, H. Garcilazo, F. Fern\'andez, and P. Gonz\'alez,
                Rep. Prog. Phys. {\bf 68}, 965 (2005).

\bibitem{Vij05a} J. Vijande, F. Fern\'andez, and A. Valcarce, 
                J. Phys. G {\bf 31}, 481 (2005).

\bibitem{Vij04} A. Valcarce, H. Garcilazo, and J. Vijande,
                Phys. Rev. C {\bf 72}, 025206 (2005).

\bibitem{Bal01} G.S. Bali,
               Phys. Rep. {\bf 343}, 1 (2001) and references therein.

\bibitem{Ade82} J.P. Ader, J.-M. Richard, and P. Taxil,
                Phys. Rev. D {\bf 25}, 2370 (1982);
                J.L. Ballot and J.-M. Richard,
                Phys. Lett. B {\bf 123}, 449 (1983);
                H.J. Lipkin,
                Phys. Lett. B {\bf 172}, 242 (1986);
                L. Heller and J.A. Tjon,
                Phys. Rev. D {\bf 32}, 755 (1985); {\it ibid} {\bf 35}, 969 (1987).

\bibitem{Vij07} J. Vijande, N. Barnea, and A. Valcarce,
                in preparation.

\bibitem{Dmi01} V. Dmitrasinovic,
                Phys. Lett. B {\bf 499}, 135 (2001);
                Phys. Rev. D {\bf 67}, 114007 (2003).

\bibitem{Mas87} K. Masutani,
                Nucl. Phys. A {\bf 468}, 593 (1987);
                M. Osamu,
                Nucl. Phys. A {\bf 505}, 655 (1989).

\bibitem{Oki05} F. Okiharu, H. Suganuma, T.T. Takahashi,
                Phys. Rev. D {\bf 72}, 014505 (2005);
                C. Alexandrou and G. Koutsou,
                Phys. Rev. D {\bf 71}, 014504 (2005);
                F. Okiharu, H. Suganuma, T.T. Takahashi, and T. Doi,
                AIP Conf. Proc. {\bf 842}, 231 (2006);
                H. Suganuma, H. Ichie, F. Okiharu, and T.T. Takahashi,
                hep-lat/0508001.

\bibitem{Vij07b} J. Vijande, A. Valcarce, and J.-M. Richard, ArXiv:0707.3996.

\bibitem{Jaf03} R. Jaffe and F. Wilczek,
                Phys. Rev. Lett. {\bf 91}, 232003 (2003);
                R.L. Jaffe
                Nucl. Phys. B (Proc. Suppl.) {\bf 142}, 343 (2005).

\bibitem{Kar06} M. Karliner and H.J. Lipkin,
                Phys.Lett. B {\bf 638}, 221 (2006).

\bibitem{CLE07} CLEO Collaboration, C. Cawlfield {\it et al.}, 
                Phys. Rev. Lett. {\bf 98}, 092002 (2007).

\bibitem{Bab05c} BaBar Collaboration, A. Aubert {\it et al.},
                Phys. Rev. D {\bf 71} 031501R (2005).

\bibitem{CDF06} CDF Collaboration, A. Abulencia {\it et al.},
                Phys. Rev. Lett. {\bf 96}, 102002 (2006);
                A. Abulencia {\it et al.}, 
                Phys. Rev. Lett. {\bf 98}, 132002 (2007);

\bibitem{Bel05b} Belle Collaboration, K. Abe {\it et al.}, hep-ex/0505038.

\bibitem{Bel06} Belle Collaboration, G. Gokhroo {\it et al.},
                Phys. Rev. Lett. {\bf 97}, 162002 (2006).

\bibitem{Set06} K.K. Seth, hep-ex/0511061.

\bibitem{COD} A.C. Doyle, {\it The sign of the four}, 1890.


\end{thebibliography}
\end{document}